\documentclass[reprint, aps, superscriptaddress, pra]{revtex4-2}
\usepackage{graphicx}
\usepackage{float}
\usepackage{amsmath}
\usepackage{amsfonts}
\usepackage{svg}
\usepackage{amssymb}
\usepackage{commath}
\usepackage{braket}
\usepackage{multirow}
\usepackage[utf8]{inputenc}
\usepackage[margin=1in]{geometry}
\usepackage{comment}
\usepackage{dcolumn}
\usepackage{bm}
\usepackage{textgreek}
\usepackage{upgreek}
\usepackage{xr-hyper}
\usepackage{hyperref}
\usepackage[capitalize]{cleveref}
\usepackage{siunitx}
\usepackage{xcolor}
\usepackage[normalem]{ulem}
\usepackage{booktabs}
\usepackage{makecell}
\usepackage{hhline}
\makeatletter
\newcommand*{\addFileDependency}[1]{%
  \typeout{(#1)}
  \@addtofilelist{#1}
  \IfFileExists{#1}{}{\typeout{No file #1.}}
}
\makeatother

\usepackage{tikz,xcolor,hyperref}

\definecolor{lime}{HTML}{A6CE39}
\DeclareRobustCommand{\orcidicon}{
	\begin{tikzpicture}
	\draw[lime, fill=lime] (0,0) 
	circle [radius=0.16] 
	node[white] {{\fontfamily{qag}\selectfont \tiny ID}};
	\draw[white, fill=white] (-0.0625,0.095) 
	circle [radius=0.007];
	\end{tikzpicture}
	\hspace{-2mm}
}
\foreach \x in {A, ..., Z}{\expandafter\xdef\csname orcid\x\endcsname{\noexpand\href{https://orcid.org/\csname orcidauthor\x\endcsname}
			{\noexpand\orcidicon}}
}
\begin{document}

\preprint{APS/123-QED}

\title{Spectral tuning of single T centres by the Stark effect}

\author{Michael Dobinson\orcidA{}}
\thanks{These two authors contributed equally}
\affiliation{Department of Physics, Simon Fraser University, Burnaby, British Columbia, Canada}
\affiliation{Photonic Inc., Coquitlam, British Columbia, Canada}
\author{Felix Hufnagel}
\thanks{These two authors contributed equally}
\affiliation{Department of Physics, Simon Fraser University, Burnaby, British Columbia, Canada}
\affiliation{Photonic Inc., Coquitlam, British Columbia, Canada}
\author{Simon A. Meynell}
\affiliation{Department of Physics, Simon Fraser University, Burnaby, British Columbia, Canada}
\affiliation{Photonic Inc., Coquitlam, British Columbia, Canada}
\author{Camille Bowness}
\affiliation{Department of Physics, Simon Fraser University, Burnaby, British Columbia, Canada}
\affiliation{Photonic Inc., Coquitlam, British Columbia, Canada}
\author{Melanie Gascoine\orcidC{}}
\affiliation{Department of Physics, Simon Fraser University, Burnaby, British Columbia, Canada}
\affiliation{Photonic Inc., Coquitlam, British Columbia, Canada}
\author{Walter Wasserman}
\affiliation{Photonic Inc., Coquitlam, British Columbia, Canada}
\author{Prasoon K. Shandilya}
\affiliation{Photonic Inc., Coquitlam, British Columbia, Canada}
\author{Christian Dangel}
\affiliation{Photonic Inc., Coquitlam, British Columbia, Canada}
\author{Michael L.W. Thewalt\orcidD{}}
\affiliation{Department of Physics, Simon Fraser University, Burnaby, British Columbia, Canada}
\affiliation{Photonic Inc., Coquitlam, British Columbia, Canada}
\author{Stephanie Simmons}
\affiliation{Department of Physics, Simon Fraser University, Burnaby, British Columbia, Canada}
\affiliation{Photonic Inc., Coquitlam, British Columbia, Canada}
\author{Daniel B. Higginbottom\orcidB{}}
\email{dhigginb@sfu.ca}
\affiliation{Department of Physics, Simon Fraser University, Burnaby, British Columbia, Canada}
\affiliation{Photonic Inc., Coquitlam, British Columbia, Canada}

\date{\today}

\begin{abstract} 
Among the many solid-state emitters being explored for scalable quantum technologies, the silicon T centre is a leading candidate offering long-lived spin qubits, a telecommunications-band spin-photon interface, and integration with on-chip photonic circuits. However, nanophotonic integration broadens both the inhomogeneous spectral distribution and individual emitter linewidths. Here, we integrate single T centres into silicon nanophotonic cavities with p-i-n diodes for local electronic control. These devices enable Stark tuning up to 30~GHz, sufficient to bring 55(2)\% of on-chip T centres into mutual resonance, and demonstrate tunable lifetime reduction across the cavity resonance. A model of the joint excitation probability shows an orders-of-magnitude increase in entanglement rate by tuning distinct emitters into mutual resonance. Luminescence modulation at high reverse biases reveals a transition to a dark charge state. Finally, bias-induced modulation of the optical transition splitting uncovers a potential mechanism for electrically driven excited-state spin mixing via spin--orbit coupling. Localized and individual spectral tuning increases the yield of performant silicon spin-photon interfaces and the number of devices per chip available for large-scale entanglement and quantum information technologies.
\end{abstract}

\maketitle

\begin{figure*}
  \makebox[\textwidth][c]{\includegraphics[width=180mm]{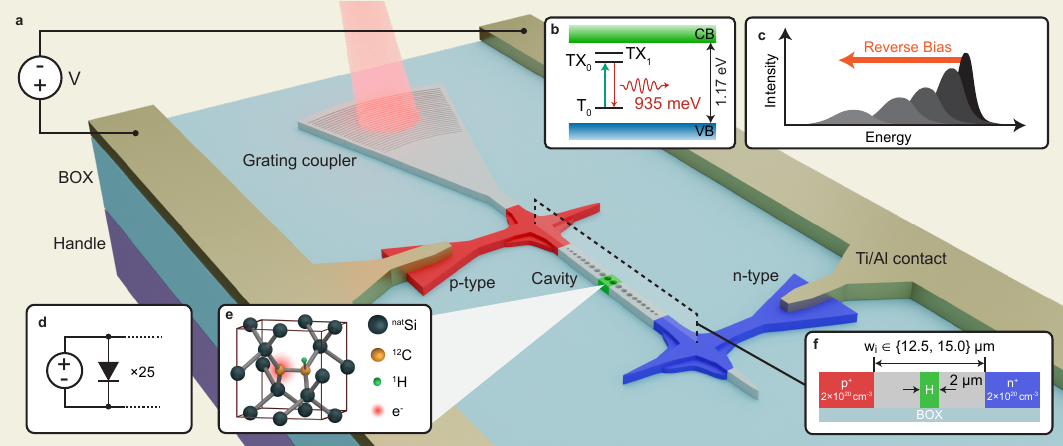}}%
  \caption{\textbf{Overview of the hybrid nanophotonic cavity and p-i-n diode architecture. a}, Illustration of the device consisting of a grating coupler, nanophotonic cavity, and p-i-n diode. The p-doped (red) and n-doped (blue) regions are connected electrically through waveguide crossings to form a diode across the nanophotonic cavity. Varying the voltage applied to the diode modulates the electric field within the device. \textbf{b}, Energy levels of the T centre within the silicon bandstructure. The ground state, T$_0$, and excited state, TX$_0$, of the T centre are shown alongside the conduction band minimum (CB) and valence band maximum (VB). \textbf{c}, Sketch of the ZPL emission spectrum illustrating the experimentally observed shifting and broadening in response to an applied reverse bias. \textbf{d}, Diagram showing the circuit for the device bus consisting of 25 cavity/p-i-n diode devices wired in parallel.. \textbf{e}, Chemical structure of the T centre within the silicon unit cell. \textbf{f}, Doping and intrinsic region width details for the p-i-n diode. Hydrogen implantation was confined to $\pm1~\upmu$m at the centre of the intrinsic region.}
  \label{fig:device_details}
\end{figure*}

\section{Introduction}
\label{sec:introduction}
Colour centres in solid-state platforms have applications in quantum computing and quantum networking~\cite{Awschalom2018, Simmons2024}. Among these, the silicon T centre has emerged as a candidate offering long-lived spins and a spin-photon interface with emission in the telecommunications O-band~\cite{Bergeron2020,Higginbottom2022}. These properties enable low-loss transmission over optical fibre and integration with mature silicon manufacturing~\cite{Simmons2024}. Nanophotonic integration can enhance the photon emission rate and coherence by the Purcell effect~\cite{Johnston2023cavity,Islam2023cavityenhanced}, but also introduces variations in the local strain and charge environment which produce an inhomogeneous spectral distribution, typically on the order of $30$~GHz~\cite{Johnston2023cavity}. Common entanglement schemes require mutual resonance between two separate emitters~\cite{Barrett2005,Kambs2018}, simultaneously resonant with two separate optical cavities. Spectral tuning techniques allow for local compensation of fabrication inhomogeneity, increasing the proportion of devices ready for entanglement at scale.  

Integrated p-i-n diodes may be employed to tune emitters by the direct-current (DC) Stark effect, a method previously demonstrated for defects in diamond~\cite{DeSantis2021}, SiC~\cite{Anderson2019, zeledon2025Minutelong, Steidl2025Single}, and the silicon G centre~\cite{day2024Electrical}. The T centre's bound-exciton optical excited state is highly sensitive to electric fields \cite{Clear2024optical,Alaerts2025}.  However, recent measurements of T centre ensembles in p-i-n diodes showed no observable Stark shift below the breakdown voltage~\cite{day2025Probing}.

Here, we demonstrate spectral tuning of single T centres in nanophotonic cavities with p-i-n diodes, introducing a tuneable, integrated, and telecommunications-band-compatible spin-photon device for integration with silicon photonics. We observe DC Stark shifts of up to 30~GHz, providing sufficient range to bring 55(2)\% of the inhomogeneous ensemble into mutual resonance. We find that spectral tuning is accompanied by spectral broadening, which we attribute to increased electric field sensitivity. To evaluate the practical trade-offs, we modelled the Hong-Ou-Mandel (HOM) visibility and joint excitation probabilities using the experimentally measured Stark tuning and broadening rates. Modelling the joint excitation probability under resonant excitation shows that tuning distinct emitters into mutual resonance can increase the joint excitation probability by over five orders of magnitude, which translates to a proportional increase in remote entanglement rates~\cite{Photonic2024}. This approach enhances the feasibility of T centre devices for scalable quantum networking and computing applications. We also observe several phenomena that illuminate the physics of the T centre including modulation of the luminescence intensity beyond an electric field threshold, consistent with conversion to a dark charge state, and modulation of the spin-dependent optical transition splitting. 

\section*{Results}
\subsection*{Device design}
The devices studied in this work (Fig.~\ref{fig:device_details}a, previously described in Ref.~\cite{dobinson2025Electrically}) utilize single T centres in a 1-D nanophotonic cavity with a p-i-n diode. The doped regions of the p-i-n diode are connected through waveguide crossings to provide electrical connection with moderate optical loss ($\sim2.2$~dB, at 295~K). A grating coupler couples an off-chip resonant laser for optical excitation and collects emission from cavity-coupled T centres. The devices were fabricated in high-resistivity 220~nm (100) silicon on insulator (SOI) using masked ion implantation of the p- and n-type dopants, with implantation parameters chosen for degenerate doping to minimize carrier freezeout. The T centres were fabricated using a process adapted from Ref.~\cite{MacQuarrie2021}, with masked hydrogen implantation confining T centre formation to $\pm1~\upmu$m from the middle of the cavity~\cite{day2024Electrical}. The p-i-n diode (Fig.~\ref{fig:device_details}f) is formed by the doped regions on either end of the nanophotonic cavity, creating an intrinsic region of 12.5~$\upmu$m (15.0~$\upmu$m) for device A (devices B and C). The diodes are electrically connected in parallel busses of 25 devices (Fig.~\ref{fig:device_details}d). Diode behaviour was simulated with Synopsys Sentaurus TCAD (Supplementary Section~\ref{supmat:pin_simulations}). The devices were cooled in a closed-cycle cryostat with measurements performed at $T=2.5$~K, unless otherwise noted. Measurements of the current-voltage (IV) characteristics show leakage currents of $<10$~nA for the busses of 25 devices over the operating range at 2.5~K (Supplementary Section~\ref{supmat:iv_curves}).

\subsection*{Spectral tuning of single T centres}
Applying a bias voltage to the p-i-n diode modulates the electric field in the intrinsic region, shifting the T centre's energy levels (see Fig.~\ref{fig:device_details}b) by the DC Stark effect. Previously, a DC Stark shift has been observed for T centre ensembles in bulk silicon~\cite{Clear2024optical}. Integrating the nanophotonic cavities in p-i-n diodes  allows the electric field to be independently controlled for separate devices on a chip, enabling individual T centres to be tuned into resonance with the cavity, or spectrally aligned for remote entanglement protocols. The device was operated under reverse bias to increase the electric field in the intrinsic region with minimal leakage current and heating.

\begin{figure*}
  \makebox[\textwidth][c]{\includegraphics[width=180mm]{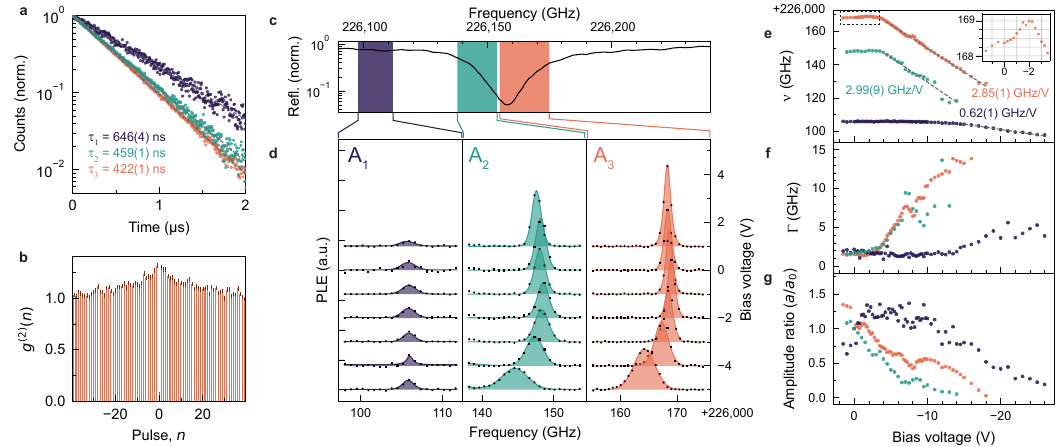}}%
  \caption{\textbf{Stark tuning of single cavity-coupled T centres. a}, Time-resolved luminescence decay transient of the zero-phonon line (ZPL) under pulsed resonant excitation at 0~V. The data are normalized and then fit to single exponential functions (grey dashed lines) to extract the excited-state lifetime. \textbf{b}, Background-corrected second-order correlation function for centre $A_3$ under pulsed resonant excitation, yielding $g^{(2)}(0)=0.090(6)$. \textbf{c}, Reflection spectrum of the cavity with a fitted Lorentzian FWHM of 41(1) GHz. \textbf{d}, PLE spectra of the ZPLs for three separate T centres ($A_1$, $A_2$, and $A_3$), demonstrating spectral shifting in response to an applied bias voltage. Solid lines indicate fits to Gaussian-Lorentzian product functions. \textbf{e-g}, Bias-dependent response of the three centres, showing the extracted (e) frequency, $\nu$, (f) linewidth, $\Gamma$, and (g) normalized amplitude ratio $a/a_0$, where $a_0$ is the amplitude at 0~V. In (e), grey dashed lines represent linear fits to the data, and the inset shows the blue-shifting region for centre $A_3$ indicated with the black dashed rectangle.}
  \label{fig:stark_shift}
\end{figure*}
Initial optical characterization of the devices was performed with the diode shorted (0~V). The cavity resonance of device A is shown in Fig.~\ref{fig:stark_shift}c, and we determine a cavity quality factor $Q=5500(100)$. Spectrally isolated single T centres were identified using pulsed photoluminescence-excitation (PLE) spectroscopy (Methods). Emission from these centres was determined to be predominantly single-photon by a pulsed Hanbury-Brown-Twiss experiment, yielding $g^{(2)}(0)=0.090(6)$ with an uncorrected $g^{(2)}_{\mathrm{raw}}(0)=0.34(1)$ for centre $A_3$ (Fig.~\ref{fig:stark_shift}b), primarily limited by the detector dark count rate. Pulsed resonant excitation and time-resolved single-photon detection was used to measure the excited state lifetime, as shown for three representative centres in Fig.~\ref{fig:stark_shift}a. We observe linewidths for single centres in the range of 1.5--5.6~GHz, which is within the reported range for integrated T centres at this temperature~\cite{DeAbreu2023waveguide,Johnston2023cavity,Islam2023cavityenhanced}, and attributed to spectral diffusion~\cite{Bowness2025,Zhang2025}.

The DC Stark effect was characterized by recording PLE spectra under applied bias (Fig.~\ref{fig:stark_shift}d). We fit the spectral peaks from multiple T centres coupled to the optical mode with Gaussian-Lorentzian product functions to extract the emission frequency shift, linewidth, and amplitude (Fig.~\ref{fig:stark_shift}e--g). We focus on three centres from device $A$ (intrinsic region width, $w_i=12.5~\upmu$m) in Fig.~\ref{fig:stark_shift}: $A_1, A_2, $ and $A_3$. Each centre exhibits a distinct response for a given bias voltage with centre $A_3$ having the widest tuning range: a red-shift of 30~GHz at $-14$~V and an initial blue-shift of $0.6$~GHz at $-2$~V. In contrast, centre $A_1$ presents a red-shift of 8~GHz at $-26$~V, with no observed blue-shifting. A 30~GHz red-shift tuning range enables $55(2)\%$ of T centres on this chip to be tuned into mutual resonance, based on the inhomogeneous distribution measured by PLE of a T centre ensemble in a waveguide device (Supplementary Section~\ref{supmat:inhomogeneous}). To account for the amplitude reduction observed in some centres at high biases (Fig.~\ref{fig:stark_shift}g), restricting the tuning range to 10~GHz yields a more conservative estimate for the tuneable fraction of $20(1)\%$.

\begin{table}[b]
    \centering
    \caption{Parameters for centres $A_1, A_2,$ and $A_3$. Centre frequency $\nu_0$ and linewidth $\Gamma_0$ at 0~V, threshold voltage $V_T$, linear tuning rate $\alpha_1$, and linear broadening rate $\gamma_1$. Note that as the device operates under reverse bias ($V<V_T\leq 0$), the negative broadening rates ($\gamma_1$) correspond to a net positive increase in the observed linewidth as the reverse bias increases.}
    \begin{tabular}{l c c c c c}
        \hline\hline
         & $\nu_0$ & $\Gamma_0$ & $V_T$ & $\alpha_1$ & $\gamma_1$\\
        No. & (GHz) & (GHz) & (V) & (GHz/V) & (GHz/V) \\
        \hline
        $A_1$ & 226,105.8(2) & 2.2(3) & $-14$ & 0.62(1) & -0.29(2)\\
        $A_2$ & 226,148.18(2) & 1.75(3) & $-4$ & 2.99(9) & -1.16(3)\\
        $A_3$ & 226,168.37(1) & 1.52(2) & $-4$ & 2.85(1) & -1.01(1)\\
        \hline\hline
    \end{tabular}
    
    \label{tab:t_params_table}
\end{table}

A voltage threshold ($V_T$) defines the region of minimal spectral tuning, indicating a small local electric field. Beyond this threshold, for centres $A_1$, $A_2$, and $A_3$, the spectral tuning is approximately linear with voltage, with the emitter linewidth increasing linearly, and the amplitude decreasing until the luminescence is fully quenched. We note that this quenching is repeatable with  steady-state recovery across multiple trials. The change in frequency $\Delta \nu$ and change in FWHM linewidth $\Delta\Gamma$ follow empirical linear relations beyond the threshold voltage:
\begin{subequations}
\begin{align}
    \Delta\nu(V)=\alpha_1(V-V_T), \;\;\;\; & V<V_T \label{eq:freq_change}\\
    \Delta\Gamma(V)=\gamma_1(V-V_T), \;\;\;\; & V<V_T \label{eq:lw_change}
\end{align}
\end{subequations}

where $\alpha_1$ and $\gamma_1$ are the linear Stark tuning and broadening rates, respectively. Table~\ref{tab:t_params_table} summarizes the threshold voltages for each centre and their respective linear tuning and broadening rates. Variations in the electric field responses between centres are expected due to the different possible orientations of the T centre~\cite{Clear2024optical} and their distinct positions within the diode (with the masked hydrogen implantation confining centres to $\pm 1~\upmu$m). A summary of all measured T centres, higher-order polynomial fits incorporating a quadratic rate, and model selection criteria are given in Supplementary Section~\ref{supmat:t_centre_table}.

\begin{figure}
  \makebox[\linewidth][c]{\includegraphics[width=86mm]{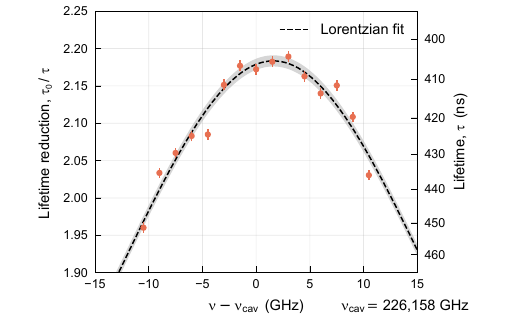}}%
  \caption{\textbf{Lifetime reduction by tuning into the cavity resonance.} The emission of centre $A_3$ is Stark tuned across the cavity resonance, with the excited-state lifetime measured at specific spectral positions by pulsed resonant optical excitation (orange circles). The detuning $\nu-\nu_{cav}$ represents the offset from the cavity resonance at $\nu_{cav}=226,158$~GHz. Error bars are centred on the datapoints and correspond to $\pm1$ standard deviation of the propagated fit error. The dashed black line indicates a Lorentzian fit to the data, with the shaded grey region representing a $\pm1$ standard deviation confidence interval of the fit.}
  \label{fig:cavity_tuning}
\end{figure}

We eliminate temperature changes due to Joule heating as a potential cause of the observed shifts. First, the dissipated electrical power remains negligible across all reverse biases used in this work (Supplementary Section~\ref{supmat:iv_curves}). Second, heating produces a uniform red-shift inconsistent with the observed behaviour~\cite{Bergeron2020}. We find that individual centres localized in the same cavity mode exhibit distinct tuning responses, including blue-shifts of up to $0.6$~GHz. Finally, we measure the low-power homogeneous linewidth of centre $A_3$ at $T=1.6$~K by saturation hole burning spectroscopy, extracting a linewidth of $\lesssim~31(5)~\mathrm{MHz}~(\lesssim37(8)~\mathrm{MHz})$~for bias voltages of $0~\mathrm{V}$~($-4~\mathrm{V}$, corresponding to a $0.9$~GHz red-shift). We find no evidence of the homogeneous linewidth broadening that is expected with thermal spectral tuning (Supplementary Section~\ref{supmat:joule_heating})~\cite{Bergeron2020}.

Figure~\ref{fig:cavity_tuning} shows the lifetime reduction of centre $A_3$ as it is tuned across the optical cavity resonance at $T=1.6$~K. We observe a modification of the excited state lifetime that follows the Lorentzian cavity profile, yielding a maximum lifetime reduction of $\tau_0/\tau=2.18(1)$ (See Methods for details). 

We measure the frequency modulation rise time for centre $A_2$ by monitoring the luminescence while applying an electrical pulse to tune the centre into resonance with the excitation laser, which is detuned from the 0~V peak position (Supplementary Section~\ref{supmat:pulsed_stark})~\cite{Cao2014Ultrafast}. We measure a rise time of $t_{10-90\%}=160$~ns, limited by the RC time constant of the device and the pulse amplifier bandwidth. Fast switching may enable schemes where individual qubits are electrically tuned into resonance with a global control field, significantly reducing the optical I/O requirements for large-scale quantum processors~\cite{Kane1998}.

\subsection*{Observations on linewidth}
Along with the observed tuning of single centres, we find that the spectral diffusion linewidth increases with applied field. This is consistent with earlier measurements of the electrical response of T centre ensembles \cite{Clear2024optical}. Spectral diffusion arises from the interaction between an emitter and  fluctuating local fields to which it is susceptible. A major contributor for T centres in integrated devices is electric field noise from fluctuating charge traps. For example, centre $A_3$ has an inhomogeneous linewidth of $1.52(2)$~GHz at 0~V, which is broader than its low-power homogeneous linewidth by over an order of magnitude (Supplementary Section~\ref{supmat:joule_heating}). An applied electric field is known to modify~\cite{Clear2024optical} the T centre's electric field sensitivity $(\frac{\mathrm{d}f}{\mathrm{d} E})$ which can thereby increase spectral diffusion.

While some studies have shown emitters with significantly reduced spectral diffusion in p-i-n diodes due to charge depletion, we observe no comparable effect in these devices. We note that diode SD reduction has been observed for emitters located deep in the substrate, far from interfaces~\cite{Anderson2019, Steidl2025Single, zeledon2025Minutelong}. In contrast, the emitters reported here are in SOI cavity devices with $\leq100$~nm to the nearest interface. Interface charge traps and surface states may not be effectively depleted by the p-i-n diodes used in this work. 

\subsection*{Intensity modulation}
\begin{figure}
  \makebox[\linewidth][c]{\includegraphics[width=86mm]{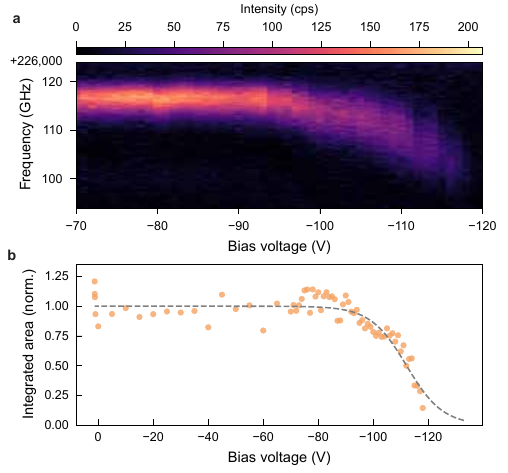}}%
  \caption{\textbf{Modulation of the ZPL intensity with applied bias. a}, The optical response of cavity-coupled T centres in a $15~\upmu$m device, showing the intensity modulation of centre $B_1$ under applied voltages from $-70$~V to $-120$~V, captured with PLE spectroscopy. \textbf{b}, The normalized integrated area of the fitted skewed Voigt function for centre $B_1$ (purple circles). The area decreases continuously past a threshold until the emission is fully quenched at $-120$~V. The integrated area data are fitted to a sigmoid activation function (black dashed line), a model relevant for intensity modulation by both field ionization and Fermi level crossing.}
  \label{fig:intensity_mod}
\end{figure}

In addition to spectral tuning, large reverse bias voltages cause repeatable luminescence quenching. We investigate this intensity modulation by considering four potential mechanisms: excess broadening, the quantum confined Stark effect, field ionization (tunnelling), and charge state conversion. To determine if this effect is caused by excess broadening, we measure the PLE peak area. Although centres $A_1 -A_3$ showed large spectral shifts, device $A$ possessed too many T centres to precisely extract the peak areas as they shifted. Therefore, we measure centre $B_1$ in a different device (Device $B$, $w_i=15.0~\upmu$m). Figure~\ref{fig:intensity_mod}a shows the response for centre $B_1$, where the luminescence is fully quenched for reverse biases $>120$~V, with the leakage current remaining low ($<10$~nA). Fitting the peak to a skewed Voigt function we find the peak area remains constant below a threshold voltage before rapidly reducing to zero (Fig.~\ref{fig:intensity_mod}b). This sharp reduction in area as the peak shifts towards the cavity resonance ($\nu_{\mathrm{cav}}=226,081$~GHz), while the peak amplitude remains above the noise floor, confirms that the luminescence modulation is not caused by excess broadening.

The quantum confined Stark effect can cause intensity modulation as the applied electric field increases the spatial separation of the electron and hole wavefunctions, reducing the oscillator strength and increasing the radiative lifetime~\cite{Aghaeimeibodi2019}. This is not compatible with the observed lifetime of centre $A_2$ showing lifetime enhancement consistent with the cavity profile and no evidence of an increased radiative lifetime as it is tuned by the electric field (Fig.~\ref{fig:cavity_tuning}). As the excited state lifetime does not increase prior to quenching we rule out a reduction in oscillator strength, characteristic of the quantum confined Stark effect, as the primary mechanism for the observed intensity modulation.

Field ionization and charge state conversion are two processes which may be responsible for the observed intensity modulation, with both causing a transition of the T centre to a dark charge state. These two processes are challenging to distinguish without precise characterization of the local electronic environment. Field ionization can occur at high reverse biases, where the electric field causes bound charges to tunnel out of the defect potential~\cite{Vincent1979Electric, Alaerts2025}. This may occur for the T centre with either the bound electron in the ground state, limiting absorption, or to the bound exciton in the excited state, increasing the non-radiative decay rate. A critical field ($F_{\mathrm{crit}}$) is required for ionization which depends on the binding energy and spatial extent of the wavefunction. For the T centre, the relevant critical field can be found for the delocalized hole~\cite{Alaerts2025}. Above the critical field, the hole dissociates from the TX$_0$ state, resulting in a negatively charged T centre~\cite{Alaerts2025}. Using the hole binding energy $E_b=35.0(1)$~meV~\cite{Bergeron2020} and the calculated effective spatial extent $a^*=35.64~\textrm{\AA}$~\cite{Alaerts2025}, the critical field required to dissociate the hole is $F_{\mathrm{crit}}\approx 10~\mathrm{MV/m}$. Device simulations of the p-i-n diode performed at 300~K (Supplementary Section~\ref{supmat:pin_simulations}) show that the maximum electric field achieved in a $15~\upmu$m device is $<7~\mathrm{MV/m}$ for a bias voltage of $-120$~V. Intensity modulation was also observed in the $12.5~\upmu$m device at bias voltage of $-20$~V, corresponding to a simulated field of $<1~\mathrm{MV/m}$.

The charge state of the T centre can be modified by controlling the Fermi level, resulting in intensity modulation as the dark charge state is preferentially populated, as observed in similar systems such as the G centre~\cite{day2024Electrical}. The T centre has a known $(0/-)$ charge transition level at $E_{(0/-)}=E_C-200~$meV~\cite{Bergeron2020}. Applying a reverse bias induces steep band-bending across the intrinsic region, which shifts the local quasi-Fermi levels. This thermodynamic shift pushes the single defect out of equilibrium and alters its steady-state charge occupation~\cite{Grimmeiss1977Deep}. While assessing the precise charge state population is challenging at cryogenic temperatures~\cite{Wolfowicz2021}, crossing above the $(0/-)$ charge transition level typically results in an increase of the population in the dark $(-)$ charge state. We have observed inter-conversion between the excited TX$_0$ state and a dark state in experiments with optical excitation followed by pulsed electrical bias, resulting in delayed luminescence (Supplementary Section~\ref{supmat:charge_state_shelving}). This behaviour is consistent with charge state conversion and is similar to exciton storage which has been observed in quantum dots~\cite{Lundstrom1999Exciton,Giroday2011Exciton}.

Field ionization and charge state conversion both depend strongly on the device behaviour and the position of the T centre within the device. Determining the relative contributions of each process would require modelling efforts which are outside the scope of this work. Figure~\ref{fig:intensity_mod}b shows the integrated luminescence intensity which is fully quenched at reverse bias voltages $>120$~V. The integrated area reduces with a sigmoidal dependence; fitting the normalized peak area $\tilde{A}$ (corresponding to the population of the neutral charge state) to a phenomenological model $\tilde{A}(V) = \left(1+\exp{((V_s-V)/\gamma)}\right)^{-1}$, yields a switching voltage $V_s=-112(1)$~V and a transition width $\gamma=6.6(7)$~V. This sigmoidal model arises naturally for charge state conversion with the neutral charge state population ($P^0$) following the Fermi-Dirac distribution, $P^0(V)\propto (1+\exp (\frac{E_{(0/-)}-E_{Fp}(V)}{k_B T}))^{-1}$, where the hole quasi-Fermi level $E_{Fp}(V)\propto V$. In the case of field ionization, the model results from the steady-state occupancy of the neutral charge state, $P^0(V)=\frac{\Gamma_c}{\Gamma_c + \Gamma_i(V)}$, given a constant capture rate $\Gamma_c$ and a voltage-dependent tunnelling rate $\Gamma_i\propto \exp({-1/F(V))}$, where $F$ is the applied electric field. With both processes yielding a sigmoidal response, the observed modulation is consistent with conversion to a dark charge state caused by either field ionization or Fermi level crossing.

\begin{figure}
  \makebox[\linewidth][c]{\includegraphics[width=86mm]{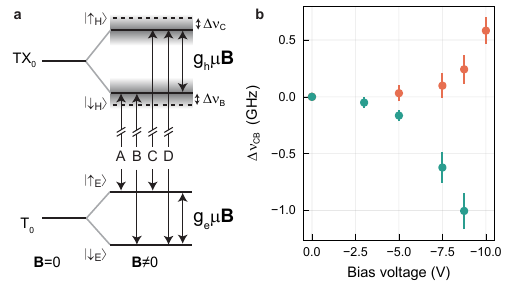}}%
  \caption{\textbf{Electronic modulation of spin-dependent optical transition splittings. a}, Level structure of the T centre under an applied magnetic field showing the four spin-dependent optical transitions, A--D. The grey shaded regions depict the shifting of the excited state energy levels due to an applied electric field. \textbf{b}, Change in splitting between the C and B transitions $\Delta \nu_{\mathrm{CB}}(V)$ for centres $A_2$ (green circles) and $A_3$ (orange circles) under an applied bias voltage with $B=600$~mT applied along the $[\bar110]$ crystal axis.}
  \label{fig:gh_efield}
\end{figure}

\begin{figure}%
  \makebox[\linewidth][c]{\includegraphics[width=86mm]{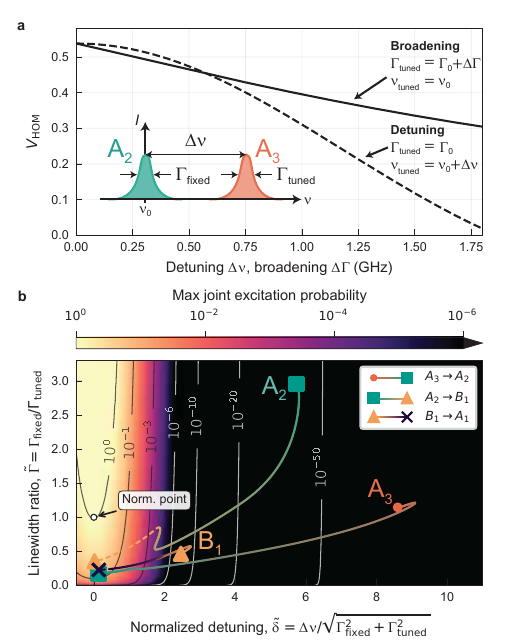}}%
  \caption{\textbf{Effects of detuning and linewidth on visibility and excitation probability. a}, Modelled HOM visibility for above-bandgap excitation of two inhomogeneously broadened quantum emitters. We model the effects of spectral diffusion (SD) and frequency detuning for emitters with lifetimes of 458~ns (corresponding to centre $A_2$) and a 0.5~ns detector gating window. The visibility is evaluated for one emitter with fixed properties while varying the second emitter in two cases: SD broadening from $\Gamma_{0}=1.52$~GHz at zero detuning (solid line), and varied detunings with fixed SD linewidths (dashed line). \textbf{b}, Modelled maximum joint excitation probability for resonant excitation of two distinct emitters. Pairs of emitters are compared using the linewidth ratio $\tilde{\Gamma}$ and normalized detuning $\tilde{\delta}$ (see Methods). Excitation probability trajectories following the experimentally observed Stark shifting and broadening are plotted for centres $A_1$ (purple cross), $A_2$ (green square), $A_3$ (orange circle), and $B_1$ (amber triangle) as they are tuned into mutual resonance, where accessible by their spectral position and tuning properties. The dashed line sections indicate the quenched luminescence regime, where the peak area is reduced by $1/e$.}
  \label{fig:SD_vis_success_prob}
\end{figure}

\subsection*{Spin-orbit coupling and splitting modulation}
In addition to the observed shifting, broadening, and intensity modulation, an applied electric field can distort the excited-state wavefunction of the T centre, modifying the orbital angular momentum character which may result in a spin-orbit interaction~\cite{Salfi2016Charge,Kobayashi2021Engineering}. This can manifest as an effective modulation of the hole $g$-factor ($g_h$) which alters the excited-state Zeeman splitting as $\Delta E =g_h \mu_B  B$ (Fig.~\ref{fig:gh_efield}a). The T centre lacks inversion symmetry ($C_{1h}$) which allows electric field perturbations to mix the heavy-hole/light-hole character ($m_j = \pm 3/2,~\pm1/2$) of the excited TX$_0$ and TX$_1$ states~\cite{Clear2024optical}, analogous to the mechanism of $g$-factor modulation observed in semiconductor quantum dots~\cite{Ares2013}. 

To investigate this, we return to device $A$ where a magnetic field of $B=600$~mT was applied along the $[\bar110]$ crystal axis. The frequency splitting between the optical C and B transitions was measured for applied biases ($\nu_{\mathrm{CB}}(V)=\nu_{\mathrm{C}}(V)-\nu_{\mathrm{B}}(V)$) on centres $A_2$ and $A_3$, chosen as they exhibited well-resolved transitions suitable for pump-probe measurements (Methods). Fully evaluating the electron and hole $g$-factors requires measuring the outer A and D transitions, however they were too weak to resolve in these measurements under reverse bias, due to the previously described intensity modulation. 

Figure~\ref{fig:gh_efield}b shows the change in the C--B splitting as a function of bias voltage for centres $A_2$ and $A_3$, $\Delta\nu_{\mathrm{CB}}(V)=\nu_{\mathrm{CB}}(V) - \nu_{\mathrm{CB}}(0)$. We observe a negative change for $A_2$ $\Delta \nu_{\mathrm{CB}}(-8.75~\mathrm{V}) =-1.0(2)$~GHz and a positive change for $A_3$ $\Delta \nu_{\mathrm{CB}}(-10.0~\mathrm{V}) = 0.58(12)$~GHz. This sign-flipping behaviour between different emitters likely arises from the distinct orientations of the defects relative to the macroscopic electric field vector, combined with variations in the local strain and charge environments. As the excited TX$_0$ state consists of a bound exciton with a tightly bound electron and a delocalized hole~\cite{Bergeron2020}, and the T$_0$ ground-state electron is also tightly bound, the electric field is expected to primarily perturb the excited-state hole wavefunction~\cite{Alaerts2025}. Consequently, we assume that the ground-state Zeeman splitting remains constant and attribute modulation of the C--B transition splitting ($\Delta\nu_{\mathrm{CB}}$) primarily to modulation of the excited-state hole $g$-factor ($g_h$).

Coupling between the hole $g$-factor and electric fields implies that charge noise can drive spin dynamics. In this manner, time-varying electric fields from nearby fluctuating charge traps may drive spin mixing in the excited state by spin--orbit coupling, consistent with reported laser-induced spin mixing for the T centre~\cite{Bowness2025}. This also suggests the potential for coherent control via electric dipole spin resonance (EDSR), a technique which has been used in quantum dots to achieve GHz-rate spin rotations~\cite{Greilich2011, Pingenot2008Method, Pingenot2011Electric, Prechtel2015Electrically}. 

\section*{Discussion}
In this work, we have demonstrated that integrating single T centres with nanophotonic devices and p-i-n diodes enables electronic control and spectral tuning at the single-emitter level. Tuning unlocks two advantages for quantum information architectures: first, it enables spectral alignment of emitters to a cavity resonance which enhances emission rates by the Purcell effect. This enhancement increases the ZPL emission efficiency by out-competing non-radiative decay pathways and phonon sideband emission~\cite{Wolfowicz2021}, increasing the efficiency beyond the intrinsic bound of $\eta_{\mathrm{QE}} \geq 23.4\%$~\cite{Johnston2023cavity}. Second, tuning separate emitters into mutual resonance improves their spectral indistinguishability, a prerequisite for high-fidelity spin-photon entanglement protocols~\cite{Barrett2005, Kambs2018}. On this chip, Stark control enables sufficient tuning range to overcome the majority of fabrication-induced inhomogeneous broadening, bringing 55(2)\% of on-chip emitters into mutual resonance. However, in these devices, tuning is accompanied by linewidth broadening, which we attribute to an increased electric field sensitivity to nearby fluctuating charge traps in regions not fully depleted by the diode. These may be un-passivated bulk, surface, or interface traps formed by implantation-related defects. Specifically, carbon and hydrogen implantation can cause hole traps in the silicon layer~\cite{Sugiyama2012} and carbon complexes that introduce positive charge at the Si/SiO$_2$ interface~\cite{Mizushima1997}. These charge traps could feasibly be minimized through, for example, localized T centre formation~\cite{Jhuria2024ProgrammableQuantum} or optimized oxide layer processing~\cite{Kim2017Annealing}. Importantly, this broadening may not be an intrinsic limitation of the tuning mechanism, but a consequence of the specific device design and fabrication methods. Future device optimization may reduce charge trap density, enabling full charge depletion while retaining a wide tuning range and maintaining the T centres in their optically active neutral charge state. 

Even without these improvements, we predict that the improved HOM visibility and excitation probability from spectral alignment outweighs the cost of SD broadening and enables interactions between previously isolated emitter pairs. We model the HOM visibility ($V_{\mathrm{HOM}}$) under above-bandgap excitation of two quantum emitters (centres $A_2$ and $A_3$) in Fig.~\ref{fig:SD_vis_success_prob}a for a 0.5~ns time-bin. Independent comparison of tuning and broadening shows that the visibility is more sensitive to uncorrected detuning than an equivalent magnitude of broadening~\cite{Kambs2018}. For a 1.5~GHz penalty, detuning degrades $V_{\mathrm{HOM}}$ for emitters $A_1$ and $A_2$ from an initial value of 0.54 to 0.12, whereas broadening preserves a visibility of 0.34. This indicates that T centres can be Stark tuned for improved photon interference, consistent with demonstrations on other platforms~\cite{Zhai2022Quantum}. Further, Fig.~\ref{fig:SD_vis_success_prob}b models the joint excitation probability for two resonantly excited emitters, which directly corresponds to the success rate of remote entanglement operations. The measured spectral positions and tuning properties of three emitters on this chip ($A_2$, $A_3$, and $B_1$) are used to calculate the joint excitation probability as they are tuned into mutual resonance with other emitters ($B_1$, $A_2$, and $A_1$, respectively). Figure~\ref{fig:SD_vis_success_prob}b visually decouples the zero-field spectral diffusion of the emitters from broadening that accompanies the tuning in the plotted trajectories, where a purely horizontal trajectory describes a tuning mechanism without additional broadening. At high reverse biases the excitation probability may be further reduced due to luminescence quenching, indicated for each tuned centre where the measured area is reduced from the initial area by a factor of $1/e$ (dashed lines). Considering the emitter pair $B_1$ and $A_1$ (with minimal luminescence quenching) indicates that tuning them into mutual resonance can increase the joint excitation probability by over five orders of magnitude. We note that while emitter $A_3$ does not show quenching across the plotted tuning range, its area is maintained due to tuning into the cavity resonance.

We report two additional noteworthy phenomena. First, at large reverse biases we observed repeatable modulation of the luminescence intensity. This behaviour is consistent with conversion to a dark charge state, driven by either field ionization or the quasi-Fermi level crossing the charge transition level. Second, we observe that an applied bias modulates the optical C--B transition splitting under a static magnetic field. We propose that this modulation may result from electric field coupling with the hole $g$-factor ($g_h$) by spin--orbit coupling, arising from an electric field-induced perturbation of the delocalized hole's orbital wavefunction. While this coupling introduces a mechanism for excited-state spin mixing driven by electrical noise~\cite{Bowness2025}, it also offers an opportunity for coherent control in the excited state. Based on the magnitude of the observed $g_h$ modulation, driving these devices with high-frequency AC electric fields presents a pathway for hole-spin EDSR~\cite{Greilich2011, Pingenot2008Method, Pingenot2011Electric, Prechtel2015Electrically}. While translating this static DC coupling to dynamic AC control requires overcoming the RC time constants of the current device architecture (Supplementary Section~\ref{supmat:pulsed_stark}), this mechanism could theoretically enable coherent and fast hole spin gates within the TX$_0$ excited state lifetime. Taken together, these demonstrations of local spectral tuning, charge state control, and electric-field-mediated spin--orbit coupling, both deepens our fundamental understanding of the T centre's electronic structure and broadens the operational toolkit for building robust and scalable quantum architectures in silicon.

\section*{Methods}
\subsection*{Device design}
The devices are fabricated on commercial high resistivity p-type (100) SOI with a 220~nm device layer~\cite{dobinson2025Electrically}. Ion implantation followed by rapid thermal annealing was used to form the p/n-doped regions as well as the T centres, with blanket carbon doping and masked implantation of P$^+$, B$^+$, and H$^+$~\cite{MacQuarrie2021,day2024Electrical}. A titanium-aluminum bilayer was used for electrical connections. An electric field was applied by the p-i-n diode (Fig.~\ref{fig:device_details}a) with intrinsic region widths of 12.5~$\upmu$m or 15~$\upmu$m, depending on the device. A grating coupler designed for the ZPL of the T centre, with a full-width at half-maximum (FWHM) bandwidth of 18(4)~meV, is used to optically couple to an $8.0^{\circ}$ angle-polished, quartz v-groove fibre array (FiberTech Optica) positioned above. Waveguide crossings are used to connect electrically to the doped region~\cite{bogaerts2007Lowloss}. Additional design information can be found in Ref.~\cite{dobinson2025Electrically}.

\subsection*{Experimental setup}
Measurements were performed in a low-vibration closed-cycle cryostat (ICEoxford $^{\mathrm{DRY}}$ICE$^{\mathrm{1.5K}}$ 85~mm) with the sample mount temperature stabilized between T=1.6--2.5~K. The sample is mounted on a gold-plated thermal link (attocube ATC100/70) on top of three-axis nanopositioner stages (attocube ANPx101 and ANPz102). The nanopositioners are used to align devices to one fibre in the six-port fibre array mounted above, routing the optical interface out of the cryostat.

A home-built pulsed laser setup was used for optical excitation. A continuously tunable laser (Toptica CTL 1320) was stabilized to a wavemeter (Bristol 871A-NIR) and swept over a spectral range. The laser power was maintained at 15~mW for mode-hop free operation, prior to the pulsing components. A high-speed pulse driver (Aerodiode SOA-std) fitted with a booster optical amplifier (Thorlabs BOA1017P) was pulsed at a current of 600~mA. The output was attenuated before passing through a synchronously pulsed electro-optic amplitude modulator
(Exail, MXER1300-LN-10) for additional extinction. Pulse timing was controlled by an ID Quantique ID900 time controller, with a digital delay generator (Stanford Research Systems DG645) to control timing offsets. The excitation pulses were routed into the device through a fibre circulator connected, outside of the cryostat, to the fibre array, and subsequently into the device's grating coupler. Emission from the device was collected back through the circulator and detected by superconducting nanowire single photon detectors (SNSPD; ID Quantique ID281), with photon arrival times recorded by the time controller.

A repetition rate of 500~kHz was used with an optical pulse duration of 100~ns for the photoluminescence excitation (PLE) measurements. The PLE spectrum was measured by sweeping the excitation wavelength and integrating the luminescence decay transient at each wavelength point for 1 second. Lifetime measurements were performed by recording the photon arrival times of the luminescence decay transient. The data were normalized and fitted with a single exponential function, $I(t)=Ae^{-t/\tau}$.

Two-colour spectroscopy was conducted using two independently-controlled tunable pulsed laser setups (as described above) which were then combined using a 50:50 fibre beam splitter before connecting to the fibre circulator which feeds the fibre array. Pump-probe measurements were performed using one laser with a fixed wavelength while the other laser was tuned, with the wavelength of the tuned laser and the integrated luminescence decay transient recorded for each point. 

Second-order correlation measurements were performed using two detectors in a Hanbury-Brown-Twiss configuration to record photon arrival times. Pulsed resonant excitation was used to excite the device and the collected emission was time gated to capture the luminescence decay transient, as described for PLE spectroscopy. Correlation measurements were captured at $T=2.5$~K ($T=1.6$~K for centres $A_1$ and $A_3$) with a pulse power of 190~nW, a pulse duration of $100$~ns, a pulse period of $4.5~\upmu$s, and a collection window of $4.25~\upmu$s. The photon arrival histograms, consisting of $d=45$~ns bins, were fitted with exponential peak functions, and their areas were used to calculate the second-order correlation for each bin. Background subtraction was performed by measuring the background count rates for each detector, $B_1$ and $B_2$, with the laser detuned from the peak. The background counts per bin were calculated by considering the cross-correlation as $\mathcal{B}=(B_1N_2+B_2N_1-B_1B_2) d T$, where $N_1$ and $N_2$ are the detector count rates and $T$ is the total measurement time. The peak area was normalized by a factor $\mathcal{N}=(N_1-B_1)(N_2-B_2)\theta T$, where $\theta$ is the pulse period~\cite{Beveratos2002}. Centres $A_1$ and $A_2$ had background-subtracted (uncorrected) $g^{(2)}(0)=0.18(2)$ and $g^{(2)}(0)=0.09(1)$ ($g^{(2)}_{\mathrm{raw}}(0)=0.61(3)$ and $g^{(2)}_{\mathrm{raw}}(0)=0.41(2)$), respectively.

\subsection*{Cavity fitting}
To extract the cavity quality factor and resonance centre, the measured spectra were fitted using a composite model that accounts for the optical resonance, linear background variation, and interference fringes. Standard lineshape models were insufficient due to the multiplicative nature of the fringes across the measurement bandwidth. The spectral data were fitted to the following phenomenological equation:

\begin{equation}
    \begin{split}
    S(\nu)= & \left[A\sin(f\nu+\phi)+B(\nu)\right]\\
            & \quad\cdot\left[B(\nu)-L(\nu;a_0,\nu_{\mathrm{cav}},\Gamma_{\mathrm{cav}})\right]
    \end{split}
\end{equation}

where $\nu$ is the frequency and the background transmission is defined as a linear function centred on the resonance: $B(\nu)=y_0+y_1(\nu-\nu_{\mathrm{cav}})$. The first term models the interference envelope with fringe amplitude $A$, spatial frequency $f$, and phase $\phi$. The second term models the underlying cavity dip where $L(\nu;a_0,\nu_{\mathrm{cav}},\Gamma_{\mathrm{cav}})$ is the Lorentzian lineshape with amplitude $a_0$, centre position $\nu_{\mathrm{cav}}$, and linewidth $\Gamma_{\mathrm{cav}}$. Sharing the background across the envelope and cavity response allows the pure Lorentzian linewidth to be extracted from the skewed background. The cavity $Q$-factor is extracted as $Q=\nu_{\mathrm{cav}}/\Gamma_{\mathrm{cav}}$.

\subsection*{Stark shift spectroscopy}
Electrical connections to the chip were made by Al wirebonds between the chip and a PCB which connected to the internal 42~SWG constantan cryostat wiring. This provided 24 external connections with an average resistance of $149(4)~\Omega$ per connection, at 295~K. The p-i-n diodes were biased by a source measurement unit capable of sourcing $\pm210$~V (Keithley 2400). Prior to optical measurements, current-voltage (IV) curves were recorded to ensure leakage currents remained negligible ($<1$~nA/device) across the operating range, avoiding Joule heating effects. Centres $A_1$ and $A_3$ were measured at $T=1.6$~K, with all other centres measured at $T=2.5$~K.

Stark shifts were extracted by recording PLE spectra at stepped DC bias voltage increments. Resonance frequencies, linewidths, and amplitudes were determined by fitting the ZPL peaks of individual T centres to a Gaussian-Lorentzian product function. To determine the integrated peak area, we fitted a skewed Voigt function using a two-step process: first, the skewness was fixed to zero to constrain the peak centre; the resulting parameters were then used to initialize a final fit where the skewness was allowed to vary. The skewed Voigt function was not used for general fitting as it proved sensitive to nearby spectral peaks.

\subsection*{Purcell factor calculation}
The lifetime reduction of a quantum emitter coupled to an optical cavity is described by the Purcell effect. We can model the lifetime reduction as a function of the detuning $\Delta\nu=\nu - \nu_{\mathrm{cav}}$ between the emitter frequency $\nu$ and the cavity resonance $\nu_{\mathrm{cav}}$ as $\tau_0/\tau_{\mathrm{cav}}(\Delta\nu)=1+\eta_{\mathrm{QE}}\eta_{\mathrm{DW}}(P_{\mathrm{cav}}(\Delta\nu)-1)$, where $\eta_{\mathrm{QE}}=0.234$ is the quantum efficiency ~\cite{Johnston2023cavity} and $\eta_{\mathrm{DW}}=0.23(1)$ is the Debye-Waller factor~\cite{Bergeron2020}. We note that this approximation assumes emission into the phonon sideband at the unmodified bulk rate. Here, the cavity Purcell enhancement is given by:

\begin{equation}
    P_{\mathrm{cav}}(\Delta\nu)=F_P\left(\frac{(\Gamma_\mathrm{cav}/2)^2}{\Delta\nu^2 + (\Gamma_\mathrm{cav}/2)^2}\right)
\end{equation}

where $\tau_0=0.885(4)~\upmu$s is the bulk lifetime under resonant excitation~\cite{Kazemi2026PRL}, $\Gamma_{\mathrm{cav}}=\nu_{\mathrm{cav}}/Q$ is the cavity linewidth (FWHM), and $F_P$ is the ideal maximum Purcell factor at resonance given by:

\begin{equation}
    F_P = \frac{3}{4\pi^2}\left(\frac{\lambda}{n}\right)^3\frac{Q}{V}
\end{equation}

where $\lambda$ is the emission wavelength (equal to the cavity centre at resonance), $n$ is the index of refraction, $Q$ is the cavity quality factor, and $V$ is the cavity mode volume. 

To experimentally evaluate the Purcell enhancement, we use the Stark effect to shift the T centre frequency across the cavity resonance. We measure the excited state lifetime at each detuning step with the laser aligned to the centre of the shifted peak. Fitting the lifetime reduction data (Fig.~\ref{fig:cavity_tuning}) to a Lorentzian profile yields an extracted cavity quality factor of $Q=4400(200)$. The amplitude of this fit corresponds to a measured effective Purcell factor of $\tilde{F_P}=23(1)$, where $\tilde{F_P} \leq F_P$ due to spatial and polarization mismatch between the emitter dipole and the cavity mode.

\subsection*{Entanglement fidelity and HOM visibility modeling} 
We model the HOM visibility $V_{\mathrm{HOM}}$ for two emitters under above-bandgap excitation with mutual detuning by a frequency $\Delta\nu$ following the Lorentzian overlap integral from Ref.~\cite{Kambs2018}. In contrast to resonant excitation, which is frequency selective (see next section), above-bandgap excitation produces photons from both emitters irrespective of $\Delta\nu$. This calculation considers a homogeneous decay rate with identical emitter lifetimes $\tau'=458$~ns and neglects pure dephasing. The HOM visibility is determined by integrating the second-order correlation function $g^{(2)}(\tau)$ for the distinguishable envelope $g^{(2)}_0(\tau)$ and an interference term $g^{(2)}_{\mathrm{int}}(\tau)$:
\begin{equation}
    \label{eq:vis_g2_difference}
    g^{(2)}(\tau)=g^{(2)}_0(\tau)-g^{(2)}_{\mathrm{int}}(\tau)
\end{equation}

The distinguishable envelope term represents the coincidence probability for distinguishable photons. For emitters with identical lifetimes, this is given by:
\begin{equation}
    g^{(2)}_0(\tau)=\frac{1}{4\tau'}\exp(-|\tau|/\tau')
\end{equation}

The interference term incorporates the detuning $\Delta\nu$ and combined spectral diffusion (SD) $\Sigma^2=\sigma_1^2+\sigma_2^2$, where $\sigma_1$ and $\sigma_2$ are the standard deviations of the Gaussian spectral diffusion profile for emitters 1 and 2, respectively. This term represents the coincidence probability due to quantum interference:
\begin{equation}
        g^{(2)}_{\mathrm{int}}(\tau) = \frac{1}{4\tau'}~\exp(-|\tau|/\tau'-2\pi^2\Sigma^2\tau^2) \cos(2\pi\Delta\nu \tau)
\end{equation}

To reduce distinguishability introduced by detuning and spectral diffusion, a time-gating window of width $\Delta \tau$ can be applied at the expense of the coincidence success rate. The interference visibility for a given window is calculated using the ratio of the integrated correlation functions:

\begin{equation}
    V_{\mathrm{HOM}}(\Delta \tau)=\frac{\int_{-\Delta \tau/2}^{\Delta \tau/2} g^{(2)}_{\mathrm{int}}(\tau)\,d\tau}{\int_{-\Delta \tau/2}^{\Delta \tau/2}g^{(2)}_0(\tau)\,d\tau}
\end{equation}

Figure~\ref{fig:SD_vis_success_prob}a evaluates the HOM visibility $V_{\mathrm{HOM}}$ for two quantum emitters under two distinct scenarios. First, we vary the SD FWHM linewidth for one emitter ($\Gamma_{\mathrm{tuned}}=\Gamma_0+f$) from the initial point $\Gamma_0=1.52$~GHz while keeping the other fixed at $\Gamma_{\mathrm{fixed}}=1.75$~GHz, assuming zero detuning. Second, we vary the detuning between emitters while maintaining identical SD linewidths. A fixed time-gating window of $\Delta \tau=0.5$~ns is used throughout.

\subsection*{Spectral overlap joint excitation probability}
We model the joint excitation probability for two inhomogeneously broadened quantum emitters (Fig.~\ref{fig:SD_vis_success_prob}b) as the maximum amplitude of the product of their Gaussian excitation profiles. One emitter is fixed at $\nu=0$ with a FWHM of $\Gamma_{\mathrm{fixed}}$ and a normalized peak amplitude of 1.0: 
\begin{equation}
    P_{\mathrm{fixed}}(\nu) = \exp\left(  \frac{-4\ln 2 ~\nu^2}{\Gamma_{\mathrm{fixed}}^2} \right)
\end{equation}

The second emitter is separated by a frequency detuning $\Delta\nu$ and has a variable FWHM of $\Gamma_{\mathrm{tuned}}$. The integrated area of the tuned emitter's excitation profile is conserved such that its amplitude scales with the broadening that accompanies tuning as $A_{\mathrm{tuned}} = \Gamma_{\mathrm{fixed}} / \Gamma_{\mathrm{tuned}}$:
\begin{equation}
    P_{\mathrm{tuned}}(\nu) = A_{\mathrm{tuned}} \exp\left(  \frac{-4\ln 2~(\nu - \Delta \nu)^2}{\Gamma_{\mathrm{tuned}}^2} \right)
\end{equation}

The product of the two Gaussians yields the joint excitation probability profile, $P_{\mathrm{joint}}(\nu)=P_{\mathrm{fixed}}(\nu)\cdot P_{\mathrm{tuned}}(\nu)$. Assuming resonant excitation where the excitation laser is aligned to the maximum of this joint profile, $P_{\mathrm{exc}}=\mathrm{max}[P_{\mathrm{joint}}(\nu)]$, which analytically evaluates to:

\begin{equation}
    \label{eq:p_exc_analytic}
    P_{\mathrm{exc}} = \left( \frac{\Gamma_{\mathrm{fixed}}}{\Gamma_{\mathrm{tuned}}} \right) \exp\left(  \frac{-4\ln 2~\Delta \nu^2}{\Gamma_{\mathrm{fixed}}^2 + \Gamma_{\mathrm{tuned}}^2} \right)
\end{equation}

For comparison between pairs of distinct emitters we reparameterize equation~\ref{eq:p_exc_analytic} by defining a normalized width ratio $\tilde{\Gamma}=\Gamma_{\mathrm{fixed}}/\Gamma_{\mathrm{tuned}}$ and a normalized detuning $\tilde{\delta}=\Delta \nu/\sqrt{\Gamma_{\mathrm{fixed}}^2 + \Gamma_{\mathrm{tuned}}^2}$. Substituting these into equation~\ref{eq:p_exc_analytic} yields a non-dimensionalized relation for the joint excitation probability:
\begin{equation}
    \tilde{P}_{\mathrm{exc}}(\tilde{\Gamma}, \tilde{\delta}) = \tilde{\Gamma} \exp\left( -4\ln 2 \cdot \tilde{\delta}^2 \right)
\end{equation}

We calculated the joint probability $\tilde{P}_{\mathrm{exc}}$ over a 2-D spatial grid, visualized using a logarithmic colourmap clipped to a lower bound of $10^{-6}$ and an upper bound of $1.0$. Tuning trajectories for emitter pairs were plotted using the experimentally determined spectral positions as well as the tuning and broadening behaviour. The trajectories are interpolated and smoothed by a univariate spline fitted to the data with a smoothing factor of 4. 

At high reverse biases luminescence modulation may further reduce the excitation probability. This regime is indicated by dashed lines in the plot where the voltage required for tuning also causes a reduction in the peak area by $1/e$. The peak area is extracted directly from the skewed Voigt fit for centre $B_1$ and is calculated by the amplitude-linewidth product for centres $A_2$ and $A_3$ which are fitted with Gaussian-Lorentzian product functions.

\section*{Data Availability}
The data that support this work are available from the corresponding author upon reasonable request.

\bibliography{references_sqt_mendeley}

\section*{Acknowledgements}
We thank the Integrated Photonics team at Photonic Inc. for their contributions to the design and fabrication of the silicon chip presented in this work. This work was supported by the Natural Sciences and Engineering Research Council of Canada (NSERC), the New Frontiers in Research Fund (NFRF), the Canada Research Chairs program (CRC), the Canada Foundation for Innovation (CFI), the B.C. Knowledge Development Fund (BCKDF), the Quantum Information Science program at the Canadian Institute for Advanced Research (CIFAR), and Photonic Inc. F.H. acknowledges support from NSERC PDF - 599491 - 2025. S.A.M. acknowledges support from NSERC Postdoctoral Fellowship 571590123. S.S. is supported by the Arthur B. McDonald Fellowship.

\section*{Author Contributions}
M.D., F.H., D.B.H., and S.S. designed the experiment. S.A.M. and C.B. theorized the spin-orbit coupling and conceived the splitting modulation experiment. M.D., F.H., S.A.M, C.B. and M.G. built the measurement apparatus. M.D. and F.H. performed the experiments. M.D performed the analysis and prepared the figures.  W.W., P.K.S, C.D. and M.L.W.T. advised on design and analysis. All authors participated in preparation of the manuscript.

\section*{Competing Interests}
M.D., F.H., S.A.M., C.B., M.G., W.W., P.K.S., C.D., M.L.W.T., S.S, and D.B.H are current or recent employees of and/or have a financial interest in Photonic Inc., a quantum technology company.

\section*{Materials \& Correspondence}
Correspondence and materials requests should be addressed to Daniel B. Higginbottom.

\onecolumngrid
\vfill
\clearpage

\renewcommand{\thefigure}{S\arabic{figure}}
\setcounter{figure}{0}
\setcounter{section}{0}
\setcounter{subsection}{0}
\vspace*{3cm}
\pagenumbering{gobble}
\begin{center}
\begin{minipage}{.8\textwidth}
\centering
\Huge{Supplementary information for ``Spectral tuning of single T centres by the Stark effect''}
\end{minipage} 

\vspace{1.5cm}

\begin{minipage}{.9\textwidth}
\centering
\Large M. Dobinson$^{1,2,*}$, F. Hufnagel$^{1,2,*}$, S. A. Meynell$^{1,2}$, C. Bowness$^{1,2}$, M. Gascoine$^{1,2}$, W. Wasserman$^{2}$, P. K. Shandilya$^{2}$, C. Dangel$^{2}$, M. L. W. Thewalt$^{1,2}$, S. Simmons$^{1,2}$, D. B. Higginbottom$^{1,2\dagger}$
\end{minipage} 

\vspace{1.5cm}

\begin{minipage}{.7\textwidth}
\centering
\large$^1$Department of Physics, Simon Fraser University, Burnaby, British Columbia, Canada \vspace{0.2cm}

$^2$Photonic Inc., Coquitlam, British Columbia, Canada \vspace{1cm}

$^{*}$These two authors contributed equally

$^{\dagger}$Corresponding author. Email: \href{mailto:dhigginb@sfu.ca}{dhigginb@sfu.ca} 
\end{minipage}

\end{center}

\normalsize
\clearpage

\section{Summary of investigated T centres}
\begin{figure}[h]
  \makebox[\linewidth][c]{\includegraphics{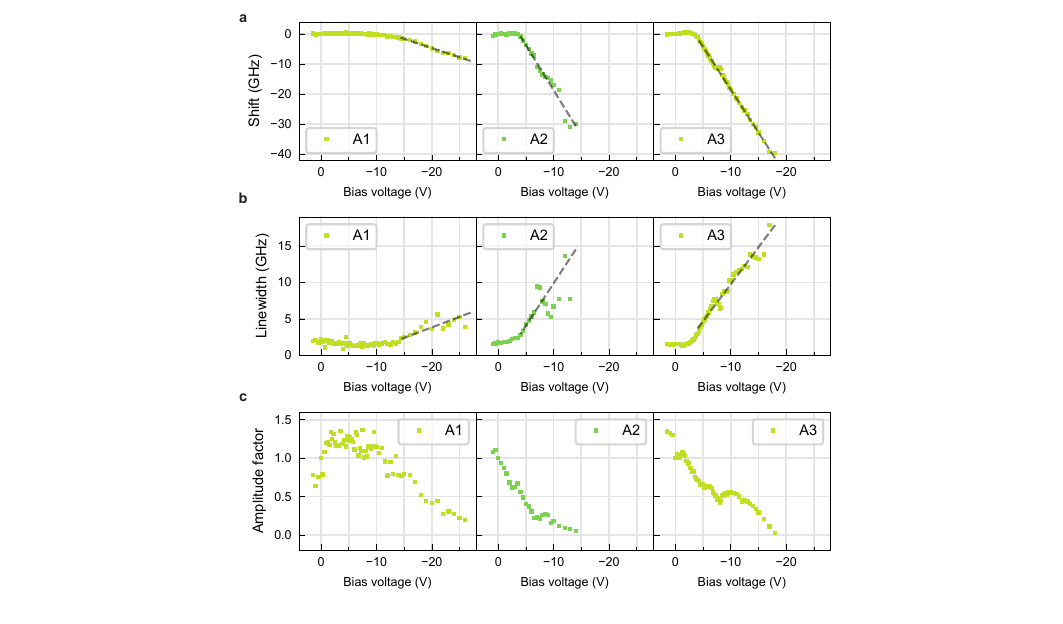}}%
  \caption{\textbf{Stark tuning characterization for T centres in the 12.5~$\upmu$m device. a}, Frequency shift ($\Delta\nu$) as a function of bias voltage. Dashed grey lines indicate the best fit to Eq.~\ref{eq:stark_shift} (linear or quadratic) over the valid tuning range. \textbf{b}, Behaviour of the ZPL transition linewidth (FWHM, $\Gamma$) with the applied bias voltage. Dashed grey lines represent fits to the broadening model in Eq.~\ref{eq:lw_broadening}. \textbf{c}, Normalized peak amplitude ($a/a_0$) vs. voltage.}
  \label{fig:stark_shift_12p5um}
\end{figure}

\begin{figure}[h]
  \makebox[\linewidth][c]{\includegraphics{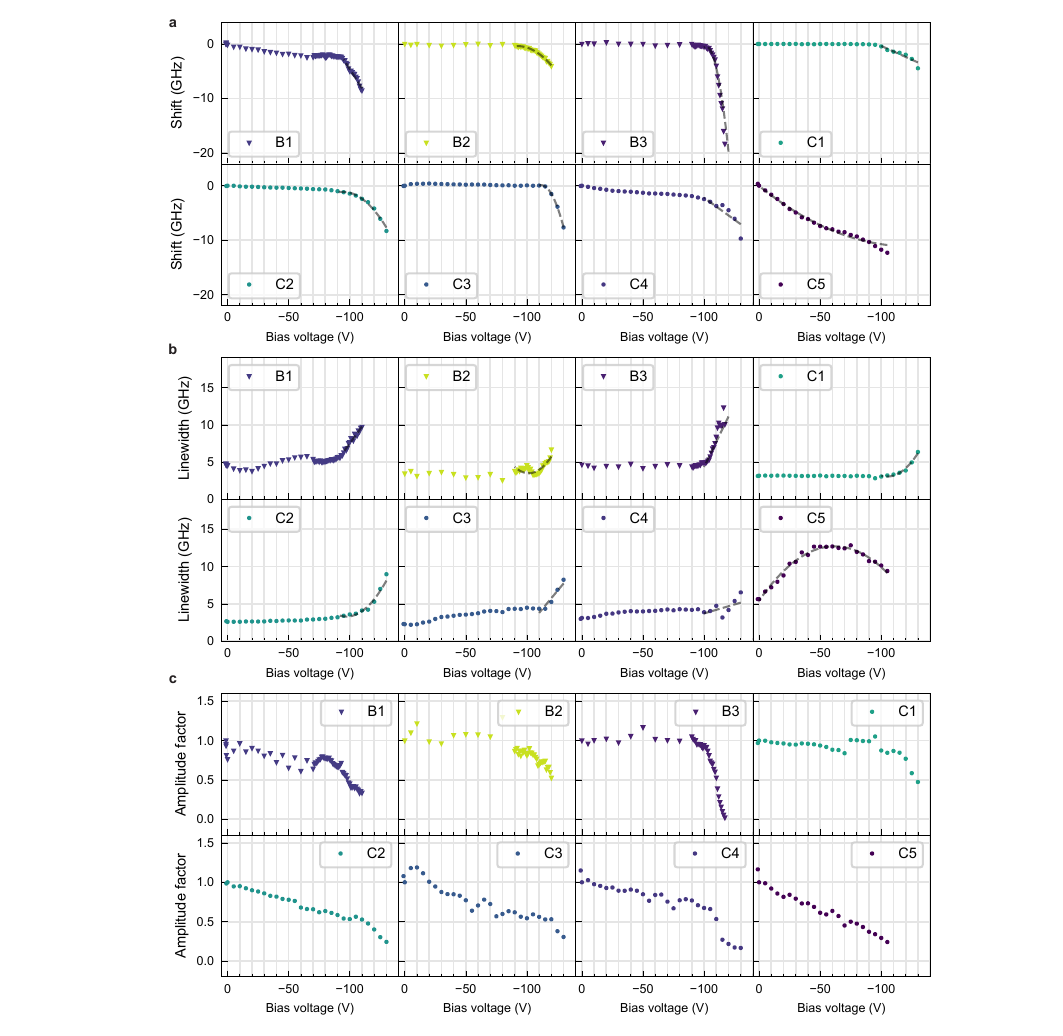}}%
  \caption{\textbf{Stark tuning characterization for T centres in 15~$\upmu$m devices. a}, Frequency shift ($\Delta\nu$) as a function of bias voltage. Dashed grey lines indicate the best fit to Eq.~\ref{eq:stark_shift} (linear or quadratic) over the valid tuning range. \textbf{b}, Behaviour of the ZPL transition linewidth (FWHM, $\Gamma$) with the applied bias voltage. Dashed grey lines represent fits to the broadening model in Eq.~\ref{eq:lw_broadening}. \textbf{c}, Normalized peak amplitude ($a/a_0$) vs. voltage.}
  \label{fig:stark_shift_15um}
\end{figure}

\pagenumbering{arabic}
\label{supmat:t_centre_table}
In this section, we provide additional details of 11 T centres investigated across three devices (A, B, and C) to support the Stark shift and linewidth broadening trends discussed in the main text. Additionally, the hole $g$-factor ($g_h$) was measured for each centre. Table~\ref{tab:T_table} summarizes the extracted parameters, while Figures~\ref{fig:stark_shift_12p5um} and ~\ref{fig:stark_shift_15um} show the individual spectral responses. As the exact electric field magnitude at the T centre is not known, we analyze the spectral response as a function of the applied voltage $V$. The dependence of the ZPL transition frequency ($\nu$) on the bias voltage ($V$) is modelled by:

\begin{equation}
    \Delta\nu(V)=\nu(V)-\nu_0 = \alpha_1 (V-V_T) + \alpha_2 (V-V_T)^2,\;\;\;V \leq V_T
    \label{eq:stark_shift}
\end{equation}

where $\nu_0$ is the zero-electric-field frequency, $V_T$ is the threshold voltage, $\alpha_1$ is the linear voltage tuning rate (related to the change in permanent dipole moment $\Delta\mu$) and $\alpha_2$ is the quadratic tuning rate (related to the change in polarizability $\Delta\alpha$). Voltage-dependent linewidth broadening is also observed, resulting from spectral diffusion due to the increased charge noise sensitivity from the applied bias. Similarly, this broadening can be described by an empirical model with linear and quadratic coefficients ($\gamma_1$, $\gamma_2$):

\begin{equation}
    \Delta\Gamma(V) = \Gamma(V) - \Gamma_0 = \gamma_1  (V-V_T) + \gamma_2 (V-V_T)^2,\;\;\;V \leq V_T
    \label{eq:lw_broadening}
\end{equation}

The frequency and linewidth responses for all T centres are listed in Table~\ref{tab:T_table}. The data were fitted with either a linear or quadratic function, selected by the Akaike information criterion (AIC) with a required improvement threshold of 5. Due to the turn-on threshold in the device response, the fit range was explicitly defined for each centre, listed in Table~\ref{tab:T_table}.

\begin{table*}[b]
  \renewcommand{\arraystretch}{1.2}%
  \centering
  \aboverulesep = 0pt
  \belowrulesep = 0pt
  \caption{\textbf{Summary of spectroscopic parameters for 11 T centres measured on devices A, B, and C.} The zero-electric-field frequency ($\nu_0$) and linewidth ($\Gamma_0$) are listed along with the Stark tuning coefficients ($\alpha_1, \alpha_2$) and voltage-dependent broadening coefficients ($\gamma_1, \gamma_2$). The parameters were extracted by fitting the data to Eq.~\ref{eq:stark_shift} and Eq.~\ref{eq:lw_broadening}. The voltage range for the fit is listed under `Fit Range' and the model (linear or quadratic) was selected using the AIC with an improvement threshold of 5, parameters for the rejected model are left blank. The hole $g$-factor ($g_h$) is included for orientation assignment. Uncertainties represent the standard error of the fit.}
  \begin{tabular}{ c r r c | r r r r | r r r r }
    \toprule
    \multicolumn{1}{c}{\multirow{3}{*}{\#}} & 
    \multicolumn{1}{c}{\multirow{3}{*}{\makecell{$w_i$ \\ \footnotesize ($\upmu$m)}}} & 
    \multicolumn{1}{c}{\multirow{3}{*}{\makecell{$g_h$}}} & 
    \multicolumn{1}{c |}{\multirow{3}{*}{\makecell{\footnotesize Fit Range \\ $[V_{min}, V_T]$\\(V)}}} &
    \multicolumn{4}{c |}{\textit{Frequency Shifting}} & 
    \multicolumn{4}{c}{\textit{Linewidth Broadening}} \\
    & & & & 
    \multicolumn{1}{c}{\multirow{2}{*}{\makecell{$\nu_0-$\footnotesize 226,000 \\ (GHz)}}} & 
    \multicolumn{1}{c}{\multirow{2}{*}{\makecell{$\Delta\nu_{\mathrm{max}}$ \\\footnotesize (GHz)}}} &
    \multicolumn{1}{c}{\multirow{2}{*}{\makecell{$\alpha_2$ \\\footnotesize (GHz/V$^2$)}}} &
    \multicolumn{1}{c |}{\multirow{2}{*}{\makecell{$\alpha_1$ \\\footnotesize (GHz/V)}}} &
    \multicolumn{1}{c}{\multirow{2}{*}{\makecell{$\Gamma_0$ \\\footnotesize (GHz)}}} &
    \multicolumn{1}{c}{\multirow{2}{*}{\makecell{$\Gamma_{\mathrm{max}}$ \\\footnotesize (GHz)}}} &
    \multicolumn{1}{c}{\multirow{2}{*}{\makecell{$\gamma_2$ \\\footnotesize (GHz/V$^2$)}}} &
    \multicolumn{1}{c}{\multirow{2}{*}{\makecell{$\gamma_1$ \\\footnotesize (GHz/V)}}} \\
    & & & & & & & & & & & \\ 
    \hhline{============}

    $A_1$        & 12.5 & 3.26(5)  & [$-27~, -14~$]  & 105.8(2)    & 8.00   & \multicolumn{1}{c}{--} & $0.62(1)  $ & $2.2(3) $ & $5.86 $ & \multicolumn{1}{c}{--} & $-0.29(2) $ \\
    $A_2$        & 12.5 & 3.01(1) & [$-14~, ~-4~$]  & 148.18(2)   & 29.87  & \multicolumn{1}{c}{--} & $2.99(9) $ & $1.75(3)$ & $14.54$ & \multicolumn{1}{c}{--} & $-1.16(3) $ \\
    $A_3$        & 12.5 & 3.26(1) & [$-18~, ~-4~$]  & 168.37(1) & 39.86  & \multicolumn{1}{c}{--} & $2.85(1) $ & $1.52(2)$ & $17.89$ & \multicolumn{1}{c}{--} & $-1.01(1)$ \\\hhline{============}
    $B_1$        & 15   & 1.21(8)  & [$-110, -95~$]  & 118.58(4)   & 5.28   & \multicolumn{1}{c}{--} & $0.35(1) $ & $4.81(8)$ & $9.77 $ & \multicolumn{1}{c}{--} & $-0.23(1) $ \\
    $B_2$        & 15   & 3.29(1)  & [$-120, -90~$]  & 113.13(7)   & 3.50   & $-0.0046(2) $ & $-0.020(6)$ & $3.5(2) $ & $5.74 $ & $0.0065(4)  $ & $0.15(1)  $ \\
    $B_3$        & 15   & 1.43(5)  & [$-120, -100$]  & 80.38(7)    & 19.08  & $-0.0580(9) $ & $-0.207(9)$ & $4.4(1) $ & $11.12$ & \multicolumn{1}{c}{--} & $-0.339(9)$ \\\hhline{============}
    $C_1$        & 15   & 2.42(4)  & [$-130, -100$]  & 115.02(3)   & 3.03   & \multicolumn{1}{c}{--} & $0.101(2) $ & $3.19(8)$ & $6.19 $ & $0.0050(5)  $ & $0.05(1)  $ \\
    $C_2$        & 15   & 2.28(5)  & [$-130, -90~$]  & 121.49(2)   & 6.47   & $-0.00515(9)$ & $-0.044(3)$ & $2.60(6)$ & $8.12 $ & $0.0045(2)  $ & $0.066(9) $ \\
    $C_3$        & 15   & 1.7(1)   & [$-130, -110$]  & 145.61(7)   & 7.67   & $-0.0232(8) $ & $-0.08(2) $ & $2.3(2) $ & $7.74 $ & \multicolumn{1}{c}{--} & $-0.20(1) $ \\
    $C_4$        & 15   & 1.41(1)  & [$-130, -100$]  & 127.07(3)   & 4.85   & \multicolumn{1}{c}{--} & $0.162(3) $ & $3.12(7)$ & $5.17 $ & \multicolumn{1}{c}{--} & $-0.048(9)$ \\
    $C_5$        & 15   & 0.95(2)  & [$-105, ~0~~$]  & 110.95(5)   & 10.84  & $0.00074(2) $ & $0.181(2) $ & $5.6(2) $ & $9.14 $ & $-0.00191(7)$ & $-0.236(7)$ \\
    \bottomrule
  \end{tabular}
  \label{tab:T_table}
\end{table*}

Figures~\ref{fig:stark_shift_12p5um} and ~\ref{fig:stark_shift_15um} show the voltage-dependent spectral response for all centres listed in Table~\ref{tab:T_table}, for devices with intrinsic region widths of $12.5~\upmu$m and $15~\upmu$m, respectively. Panel (a) in each figure displays the frequency shift $\Delta\nu$ as a function of bias voltage. The dashed grey lines show the best fit (linear or quadratic) as chosen by the AIC selection criteria. Panel (b) shows the linewidth behaviour, illustrating the voltage-dependent broadening. Panel (c) shows the normalized peak emission amplitude $a/a_0$, where $a_0$ is the amplitude at $0~V$.

\clearpage
\section{IV Curves}
\label{supmat:iv_curves}
\begin{figure}[h]
  \makebox[\linewidth][c]{\includegraphics{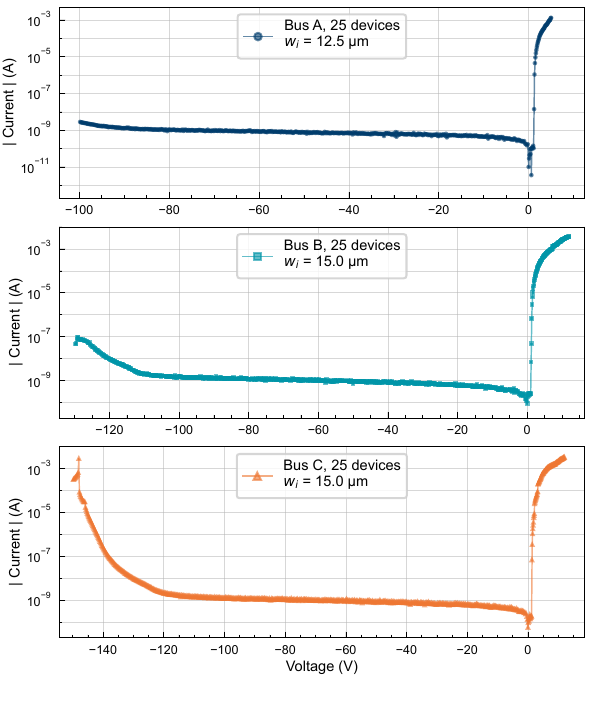}}%
  \caption{\textbf{IV curves for busses A, B, and C.} Each IV curve was taken for the entire bus of 25 devices. Bus A consists of devices with intrinsic region widths of $w_i=12.5~\upmu$m while B and C have $w_i=15.0~\upmu$m. Measurements were performed at $T=2.5$~K.}
  \label{fig:iv_12p5_and_15_um}
\end{figure}

The current-voltage (IV) characteristics of three busses of p-i-n diodes with 12.5~$\upmu$m and 15~$\upmu$m intrinsic region widths were measured at $T=2.5$~K (Fig.~\ref{fig:iv_12p5_and_15_um}) . For a reverse-bias current threshold of 1~nA, busses A, B, and C were below this threshold up to bias voltages of $-80, -60,$ and $-70$~V, respectively. A threshold current of 10~nA was also considered for busses B and C where they remained below up to bias voltages of $-120$ and $-130$~V, respectively.

\clearpage
\section{Inhomogeneous distribution}
\label{supmat:inhomogeneous}
\begin{figure}[h]
  \makebox[\linewidth][c]{\includegraphics[width=180mm]{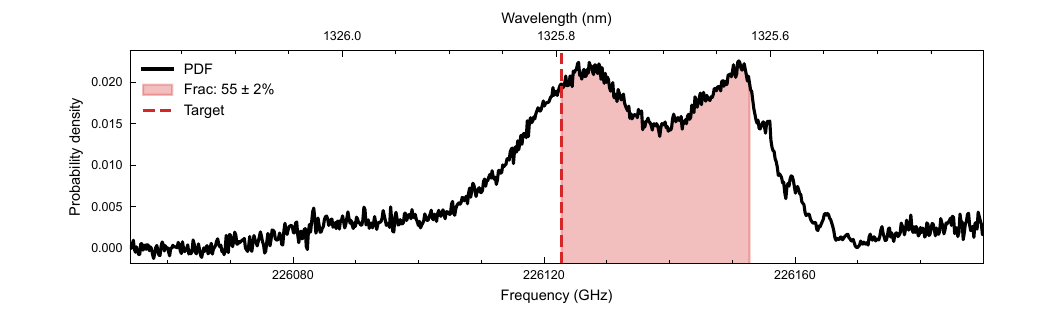}}%
  \caption{\textbf{Inhomogeneous broadening and tuning yield for a T centre ensemble.} Normalized PLE spectrum showing the inhomogeneous T centre distribution in a waveguide device with a forward bias of 1.21~V. The shaded areas within the measured probability density (black lines) indicate emitters capable of reaching a target resonance (vertical dashed line). With tuning restricted to red-shifting, 55(2)\% of emitters can be tuned to meet the target.}
  \label{fig:inhomog_dist}
\end{figure}

To determine the fraction of T centres that could be tuned into mutual resonance we first measure the inhomogeneous distribution of a T centre ensemble using PLE in a device which consists of a tapered waveguide, grating couplers, and a p-i-n diode with an intrinsic region width of  $w_i=2.5~\upmu$m (Fig.~\ref{fig:inhomog_dist}). See Ref.~\cite{dobinson2025Electrically} for additional device design details. To calculate the tuneable fraction, we employ an empirical sliding window integration. The raw intensity data, $I(\nu)$, is first background-subtracted to remove detector dark counts and baseline noise, yielding $I'(\nu)$. The data is then normalized by its total integral to generate a probability density function (PDF):
\begin{equation}
    \mathrm{PDF}(\nu) = \frac{I'(\nu)}{\int_{-\infty}^{\infty} I'(\nu) d\nu}
\end{equation}

To find the maximum tunable fraction $N_{\mathrm{tuneable}}/N$ for a tuning range $\Delta\nu$, the PDF is numerically integrated across a sliding frequency window of width $\Delta\nu$, maximizing the captured area:
\begin{equation}
    \frac{N_{\mathrm{tuneable}}}{N} = \max_{\nu_0} \int_{\nu_{0}}^{\nu_{0} + \Delta\nu} \mathrm{PDF}(\nu) d\nu
\end{equation}

where $N_{\mathrm{tuneable}}$ is the number of T centres that can be tuned into resonance with another, $N$ is the total number of T centres, $\nu_0$ is the target frequency for tuning, and $\Delta\nu$ is the tuning range. For a tuning range $\Delta\nu=30$~GHz, the tuneable fraction is $N_{\mathrm{tuneable}}/N=0.55(2)$.

\clearpage
\section{Impact of nearby fluctuating charges}
\label{supmat:fluctuating_charges}

\begin{figure*}[h]
  \makebox[\linewidth][c]{\includegraphics[width=90mm]{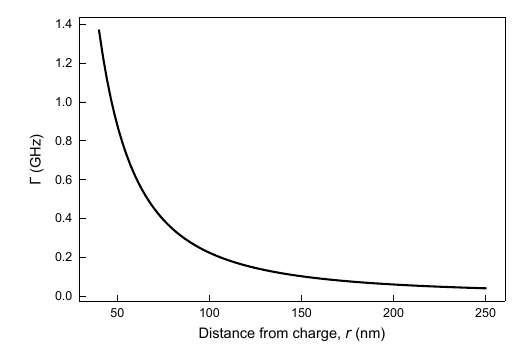}}%
  \caption{\textbf{Spectral diffusion from a fluctuating charge trap.} The inhomogeneously broadened linewidth ($\Gamma$) is plotted against the distance ($r$) from a single fluctuating charge trap. The charge is placed along the line of maximal linear coupling.}
  \label{fig:linewidth_field}
\end{figure*}

Fluctuating charge traps create a time-varying electric field which in turn can cause spectral diffusion of the T centre as the noise is coupled by the DC Stark effect. We can model the linewidth of the emission as a Lorentzian line wandering over a Gaussian profile, resulting in a Voigt lineshape~\cite{DeSantis2021}:

\begin{equation}
    \Gamma = \dfrac{\Gamma_1}{2}+\sqrt{\Big(\dfrac{\Gamma_1}{2}\Big)^2+8\ln2 [F_{\mathrm{rms}}\Delta\mu(F_{\mathrm{dc}})]^2}
\end{equation}

where $\Gamma$ is the observed linewidth, $\Gamma_1$ is the minimum linewidth, $F_{\mathrm{rms}}$ is the field from a fluctuating charge trap, and $\Delta\mu(F_{\mathrm{dc}})$ is the change in dipole moment. We can model the effect of fluctuating charge traps on the T centre linewidth using the maximal linear electric field coupling of $\Delta\mu=7519$~Hz.m/V~\cite{Clear2024optical}. The electric field from a fluctuating charge trap at a distance $r$ can be calculated as $F_{\mathrm{rms}} = \frac{1}{4\pi\epsilon_0 \epsilon_r} \frac{Q}{r^2}$. Figure~\ref{fig:linewidth_field} shows the broadening caused by a fluctuating charge placed a distance $r$ along the maximal coupling axis. In typical devices, T centres within the nanophotonic waveguide are never further than 110~nm from interfaces where charge traps are commonly present. When coupled to a cavity this distance can be much smaller and depends on the design of the cavity. Multiple fluctuating charge traps may be present at one or more nearby interfaces, causing additional broadening. Considering this, the combined impact of nearby charge traps can have a significant effect on the linewidth even if they are not maximally coupled.

\clearpage

\section{Simulation of the p-i-n diode}
\label{supmat:pin_simulations}
To investigate the electronic environment of the T centre within the p-i-n diode we performed 2-D Technology Computer-Aided Design (TCAD) process and device simulations using Synopsys Sentaurus (v. 2025.06). The simulation flow follows the fabrication steps to simulate realistic dopant distributions using a 220~nm SOI device layer on buried oxide (BOX). An adaptive mesh was used with 10~nm spacing near the interface and lateral spacing of 50~nm near the doped junctions to resolve the concentration gradients.

Accurate modelling of the electric field vector within the p-i-n diodes used in this work is challenging due to the cryogenic operating temperature and the device geometry. Nanofabrication can introduce charge traps hosted at the rough surfaces and many Si/SiO$_2$ interfaces, inherent to integration with nanophotonic waveguides and cavities. Lattice damage can be caused by the carbon/hydrogen implantation and subsequent segregation~\cite{Johnston2023cavity} can form a positively charged layer of interfacial defects~\cite{Mizushima1997}. These defects act as electronic traps which can affect the electric field direction and lower the breakdown voltage. These simulations were performed at $T=300$~K and aim to identify qualitative behaviour, in particular the rotation of the electric field vector, to guide the interpretation of the experimentally observed Stark tuning.

\subsection{Process Simulation (Sentaurus Process)}
The fabrication flow was modelled in SPROCESS to capture the dopant distributions and defect profiles resulting from implantation of the dopants and anneal cycles. Table~\ref{tab:sprocess_params} shows the relevant simulation parameters, all implantation was performed with a $7^\circ$ tilt angle.

\begin{table}[h!]
    \centering
    \caption{Summary of Sentaurus SPROCESS simulation parameters.}
    \label{tab:sprocess_params}
    \begin{tabular}{l l l}
        \hline
        \textbf{Category} & \textbf{Parameter} & \textbf{Value} \\
        \hline
        \textbf{Geometry} & Device Layer / BOX Thickness & $0.22 \, \mu\text{m}$ / $3.0 \, \mu\text{m}$ \\
                          & Intrinsic Region Width & $12.5, 15.0 \, \mu\text{m}$ \\
        \hline
        \textbf{Implantation} & \textit{p-type} (B) & $1.5 \times 10^{15} \, \text{cm}^{-2}$, 30 keV \\
                              & & (Background:  $10^{12}~\text{cm}^{-3}$)\\
                              & \textit{n-type} (P) & $1.5 \times 10^{15} \, \text{cm}^{-2}$, 62 keV \\
                              & \textit{T centre dopant} (C) & $7.0 \times 10^{12} \, \text{cm}^{-2}$, 38 keV \\
                              & \textit{T centre dopant} (H) & $7.0 \times 10^{12} \, \text{cm}^{-2}$, 9 keV \\
        \hline
        \textbf{Annealing} & Dopant Activation & $1000^\circ\text{C}$, 20~s \\
                           & Hydrogen Diffusion & $500^\circ\text{C}$, 180~s \\
        \hline
    \end{tabular}
\end{table}

\subsection{Device Physics and Defect Modelling (Sentaurus Device)}
The mesh and dopant distributions from SPROCESS were imported into SDEVICE to solve the coupled Poisson and Drift-Diffusion equations (Table~\ref{tab:sdevice_models}). To account for implantation damage we included a uniform density of background acceptor-level traps, approximating $C_iO_i$ defects~\cite{Sugiyama2012}.

\begin{table}[h!]
    \centering
    \caption{Summary of Sentaurus SDEVICE simulation parameters.}
    \label{tab:sdevice_models}
    \begin{tabular}{l p{4cm} p{8cm}}
        \hline
        \textbf{Section} & \textbf{Parameter} & \textbf{Description} \\
        \hline
        \textbf{Mobility} & DopingDep & Mobility degradation due to impurity scattering. \\
                          & HighFieldSaturation & Carrier velocity saturation at high electric fields. \\
                          & Enormal & Mobility degradation due to transverse electric fields. \\
                          & CarrierCarrierScattering & Mobility degradation due to carrier-carrier scattering. \\
        \hline
        \textbf{Recombination} & SRH (DopingDep) & Shockley-Read-Hall recombination with doping-dependent lifetimes. \\
                               & Auger & Auger recombination for high carrier densities. \\
                               & Avalanche (Okuto) & Okuto-Crowell model for avalanche generation\\
        \hline
        \textbf{Electrostatics} & EffectiveIntrinsicDensity (OldSlotboom) & Bandgap narrowing model. \\
        \hline
        \textbf{Traps}  & Acceptor Level & $C_iO_i$-related: $c=10^{15}~\mathrm{cm}^{-3}$, $E_0=E_V+0.35$~eV, $\sigma_{e}=\sigma_{h}=10^{-15}~\mathrm{cm}^{2}$\\
                        & Donor Level    & Thermal donors: $c=10^{15}~\mathrm{cm}^{-3}$, $E_0=E_C-0.035$~eV, $\sigma_{e}=\sigma_{h}=10^{-14}~\mathrm{cm}^{2}$\\
        \hline
    \end{tabular}
\end{table}

\begin{figure}
  \makebox[\linewidth][c]{\includegraphics[width=180mm]{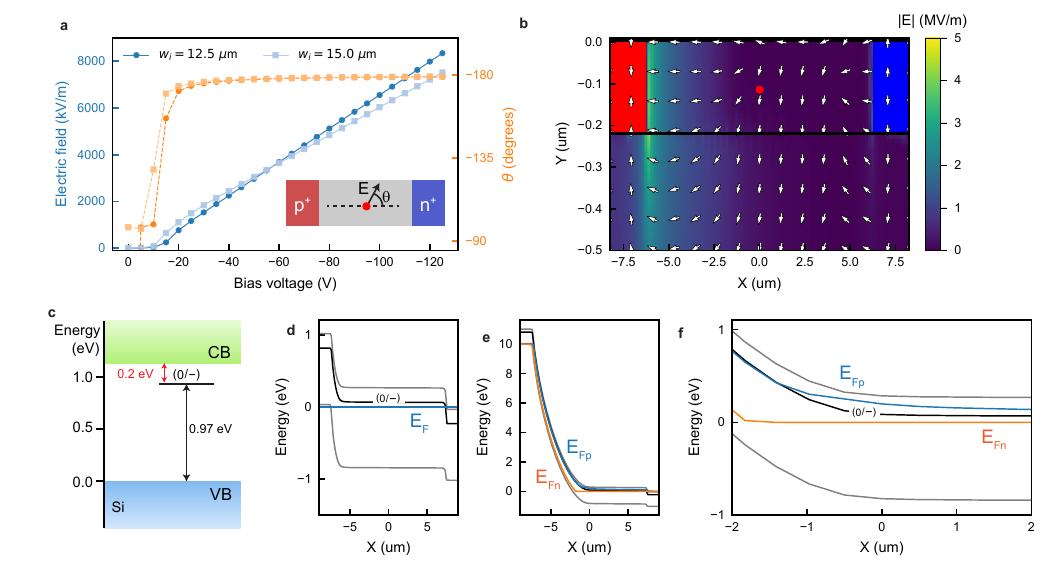}}%
  \caption{\textbf{Simulated electrostatics and band structure of the p-i-n diode. a}, Electric field amplitude and direction for a p-i-n diode with 12.5~$\upmu$m (circles) and 15.0~$\upmu$m (squares) intrinsic region. The field rotates from vertical ($-y$) to lateral ($-x$) as bias increases. Inset: Definition of the electric field vector angle $\theta$. \textbf{b}, 2-D simulation of the electric field distribution in the 12.5~$\upmu$m intrinsic region at a bias of -5~V. The colour scale indicates the electric field magnitude and the arrows depict the field vector direction. \textbf{c}, Schematic of the charge transition levels of the silicon T centre, illustrating the $(0/-)$ charge transition level relative to the band edges. Numerical values are based on experimentally determined values~\cite{Bergeron2020}. \textbf{d,e}, 1-D band diagrams for the 15.0~$\upmu$m intrinsic region device at $y=-0.11~\upmu$m for 0~V (d) and $-10$~V (e). Solid grey lines: band edges; black line: $(0/-)$ charge transition level; coloured lines: electron ($E_{Fn}$) and hole ($E_{Fp}$) quasi-Fermi levels. \textbf{f}, Magnified view of the T centre region at $-10$~V showing the transition level crossing the quasi-Fermi level for the 15.0~$\upmu$m intrinsic region device.}
  \label{fig:efield}
\end{figure}

\subsection{Electrostatic environment and charge stability}
Modelling the device electrostatics with 2-D TCAD simulations aids in interpreting the behaviour of T centres within the p-i-n diode devices. Unlike ideal p-i-n diodes, integration with the nanophotonic waveguide and cavity introduces complex field geometries. Figure~\ref{fig:efield} a and b shows the simulated electric field distribution in the intrinsic region. For low reverse biases ($|V|<10$~V) the field direction is predominantly vertical ($-y$). With increasing reverse bias, the contribution from the junction potential begins to dominate, causing the electric field vector to rotate towards the $-x$ direction.

Rotation of the electric field vector has implications for the T centre Stark shift which has an anisotropic response. The frequency shift depends on the projection of the local electric field $\mathbf{F}$ on the T centre's permanent electric dipole moment difference $\Delta\mu$. As $\mathbf{F}$ changes direction with the applied bias, the observed Stark tuning rate depends both on the field magnitude and the field angle $\theta$. This suggests that the effective tuning rate is not only dependent on the orientation of the T centre studied, but also the position of the centre relative to the nanostructure geometry, and the electronic environment.

The charge-state stability of the T centre was also investigated with these simulations for a device with an intrinsic region width of $15~\upmu$m. The T centre is known to possess a $(0/-)$ charge transition level within the bandgap (Fig.~\ref{fig:efield}c)~\cite{Bergeron2020}. Figure~\ref{fig:efield}d-f shows the 1-D band diagrams from the centre of the waveguide ($y=-0.11~\upmu$m). The charge transitions of the T centre were not included in the Poisson solver but are overlaid on the band diagrams to indicate regions of charge stability. With increasing reverse bias, band bending modifies the position of the quasi-Fermi levels ($E_{Fn}, E_{Fp}$) relative to the defect levels. At high reverse biases (Fig.~\ref{fig:efield}f) the hole quasi-Fermi level $E_{Fp}$ crosses the $(0/-)$ transition level. This crossing indicates conversion from the optically active neutral charge state (T$^0$) to the dark negative charge state (T$^-$). This provides a mechanism for the observed intensity modulation at high reverse biases, as described in the main text.

\subsection{Validity of simulations}
While the experimental characterization of the T centres in the p-i-n diodes was performed at cryogenic temperatures, the TCAD simulations presented here were conducted at 300~K to avoid numerical instabilities associated with carrier freeze-out in drift-diffusion models. This approach is used as the electrostatic profile of the p-i-n diode is governed primarily by the distribution of fixed space charges (ionized dopants and interfacial charges). As mobile carriers are essentially absent in the depletion region, the field geometry is largely determined by the Poisson equation. By simulating at 300~K, we capture the influence of the interfacial charge on electric field direction. These geometric effects and their deviation from the 1-D field are expected to persist at low temperatures, providing a qualitative guide for interpreting the Stark shift observed for different T centres.

The 2-D model shows that as the reverse bias increases, the contribution of the junction potentials increases relative to the fixed charges, causing the electric field vector to rotate (Fig.~\ref{fig:efield}a,b). As the T centre Stark shift is anisotropic, the rotation of the electric field vector is coupled to the Stark tuning rate, making precise determination of the dipole moment and polarizability challenging. We note that these simulations were performed on an idealized geometry with a single oxide interface between the device layer and the BOX. The fabricated device has potential for a more complex field profile due to contributions from charge traps hosted at the rough surfaces and additional interfaces introduced by fabrication. Additionally, band bending can modify the population of the T centre charge states (Fig.~\ref{fig:efield}d-f) which provides a mechanism for the observed bias-dependent intensity modulation.

\clearpage
\section{Joule heating}
\label{supmat:joule_heating}
To confirm that the observed frequency tuning and linewidth broadening arise from the DC Stark effect rather than local temperature variation we perform a thermal analysis  combining theoretical modelling with spectroscopic measurements. The first section estimates the device temperature rise using a heat transfer model that accounts for thermal conduction through the silicon substrate as well as conduction and convention by the surrounding helium gas. This model considers both the diffuse continuum convection and ballistic conduction regimes for the heat transfer through the helium gas. For a dissipated power of 100~nW over all devices we calculate a negligible per-device temperature rise of $\sim 0.1$~K.

The second section presents saturation hole-burning measurements to probe the local temperature of the T centre under an applied bias. Comparing the hole linewidths of Stark-shifted peaks and un-shifted peaks we find no significant difference in the homogeneous linewidth, indicating minimal thermal variation. Finally, we compare the observed spectral response to the established bulk thermal scaling laws for silicon T centres. We show that a thermal origin for the observed 0.9~GHz red-shift (at $-4$~V) would require a thermal rise of 5.7~K, inducing 0.78~GHz of homogeneous broadening. As the saturation hole-burning measurements show hole linewidths consistently below 200~MHz, this discrepancy rules out Joule heating as the mechanism for the observed spectral tuning.

\subsection{Thermal transport modelling}
The dissipated electrical power is $<100$~nW across the operating range. This power is dissipated across the bus of 25 devices as well as in the cryostat wiring. We can compute the temperature rise considering the worst case where all of the power is dissipated in a single device. Given the geometry of the waveguide with a cross-sectional area of $\sigma=0.09~\upmu\mathrm{m^2}$ and a length of $L=30~\upmu$m. We can consider two cooling scenarios: solid conduction through the silicon alone as well as conduction to and convection by the He exchange gas.

\subsubsection{Solid conduction only}
In the first scenario, heat flows only through the silicon to the heat sink.  For square rod with side lengths $D\approx\sqrt{\sigma}=0.3~\upmu$m the phonon mean free path can be approximated as $D_{\mathrm{eff}}\approx1.115D=0.3345~\upmu$m~\cite{Maris2012}. The specific heat capacity can be found using the Debye model:
\begin{equation}
    C_v=\frac{12\pi^4}{5} N_{\mathrm{Si}} k_B \left(\frac{T}{T_{D}}\right)^3 \approx 9.4~\mathrm{J}/(\mathrm{m^3}\cdot \mathrm{K})
\end{equation}

where $N_{\mathrm{Si}}\approx5\times 10^{28}~\mathrm{m^{-3}}$ is the atomic density, $k_B=1.38\times 10^{-23}~\mathrm{J/K}$ is the Boltzmann constant, $T_{D}=645$~K is the Debye temperature~\cite{Barron2016} and $T=2.5$~K is the thermal sink temperature. We can consider the nanoscopic thermal conductivity for Si in the Casimir regime as $\kappa_{\mathrm{Si}} \approx (1/3)C_v v_s D_{\mathrm{eff}} \approx 0.0067~\mathrm{W}/(\mathrm{m} \cdot \mathrm{K})$, where $v_s\approx6400\mathrm{m/s}$ is the sound velocity in silicon~\cite{Holland1963}.

The thermal resistance can be found by $R_{\mathrm{th,solid}} = L_{\mathrm{eff}}/(\kappa_{\mathrm{Si}}~\sigma)\approx 2.49\times10^{10}~\mathrm{K/W}$, where $L_{\mathrm{eff}}\approx L/2$ is the distance from the heat source to the thermal sink and $\sigma$ is the cross-sectional area. In this scenario the calculated temperature rise for a power of $P_{\mathrm{device}}\approx 4~\mathrm{nW}$ dissipated in a single device is $\Delta T_{\mathrm{solid}}\approx P_{\mathrm{device}} R_{\mathrm{th,solid}}\approx 99~\mathrm{K}$.

\subsubsection{Conduction and convection by He exchange gas}
If we consider cooling by the He exchange gas we find that heat is removed from the exterior surfaces of the device by the gas. To find the heat transfer coefficient we must first compute the Knudsen number to determine the flow regime using the mean free path $\Lambda_{c}$ and the characteristic dimension $L_c$ of the device: $\mathrm{Kn} = \Lambda_{c}/L_c$. The mean free path for an ideal gas ($\Lambda_{c}$) can be calculated independent of the gas viscosity by evaluating the typical equation~\cite{Barron2016} considering the effective cross-sectional area for spherical particles as $\pi d^2$:

\begin{equation}
    \Lambda_{C} = \frac{\eta}{p}\left(\frac{\pi R T}{2}\right)^{1/2}=\frac{k_B T}{\sqrt{2}\pi d^2 p}=66~\mathrm{nm}
\end{equation}

where $T=2.5$~K is the temperature, $d\approx 2.55~\text{\AA}$ is the effective kinetic diameter of an He atom, and $p=1800$~Pa is the gas pressure. Using the calculated mean free path and the characteristic length $L_c\approx\sqrt{\sigma}=0.3~\upmu$m we find a Knudsen number corresponding to the transition regime, i.e. $0.01<\mathrm{Kn}\approx 0.22 < 10$. This regime requires consideration of contributions from the ballistic free-molecular $h_{\mathrm{fm}}$ and continuum convection $h_{\mathrm{cont}}$ mechanisms to calculate the total effective thermal resistance. \\

\textbf{Ballistic free-molecular conduction: }We can find the heat transfer coefficient for ballistic conduction in the free-molecular regime as:

\begin{equation}
    h_{\mathrm{fm}} \approx \alpha \frac{\gamma + 1}{\gamma - 1}  \sqrt{\frac{R}{8\pi M T}}~p\approx 20,699~\mathrm{W/(m^2 \cdot K)}
\end{equation}

where $\alpha=0.5$ is a conservative estimate of thermal accommodation coefficient, $\gamma=5/3$ is the heat capacity ratio for monoatomic gases, and $\mathrm{M}\approx 4~\mathrm{g/mol}$ is the atomic mass of He. Recent work has found that the thermal accomodation constant for He on crystalline silicon can be significantly higher, up to 0.7 at 11~K which would only improve the thermal conduction~\cite{Franke2024}. \\

\textbf{Diffuse continuum convection heat transfer: }For diffusive heat transfer in the continuum convection regime, the thermal conductivity $\kappa_{\mathrm{gas}}$ of the He gas can then be found as: $\kappa_{\mathrm{gas}}\approx 15\mathrm{R}\eta/(4\mathrm{M}) = 0.0061~\mathrm{W/(m\cdot K)}$, where $\mathrm{R}=8.314 \mathrm{J/(mol\cdot K)}$ is the gas constant and $\eta\approx0.78\times10^{-6} ~\mathrm{Pa\cdot s}$ is the viscosity (calculated from the 3.5~K reference $\eta_{3.5\mathrm{K}}=0.9825\times10^{-6} ~\mathrm{Pa\cdot s}$ from Ref.~\cite{Arp1998} and the scaling factor $\eta\propto T^{0.647}$ from Ref.~\cite{Chapman1991}). The continuum heat transfer coefficient can be found by considering the waveguide as a small compared to the surrounding volume, such that the heat conduction is geometrically enhanced by the $1/r$ dependence. We approximate the waveguide as a half-cylinder in the larger cryostat volume which contains the gas:

\begin{equation}
    h_{\mathrm{cont}}\approx\frac{1}{2}\left(\frac{\kappa_{\mathrm{gas}}}{r_{\mathrm{in}} \ln{(r_{\mathrm{out}}/r_{\mathrm{in}})}}\right) \approx 1619~\mathrm{W/(m^2 \cdot K)}
\end{equation}

where $r_{\mathrm{in}}\approx L_c/2 = 0.15~\upmu$m is the approximated waveguide radius and $r_{\mathrm{out}}=42.5$~mm is the radius of the cryostat interior ($D_i=85$~mm). The total effective heat transfer coefficient can be found by the inverse sum of the free-molecular and continuum thermal resistances: $h_{\mathrm{eff}} = (h_{\mathrm{fm}}^{-1}+h_{\mathrm{cont}}^{-1})^{-1}\approx 1501~\mathrm{W/(m^2 \cdot K)}$. This results in a much lower thermal resistance which dominates the behaviour. We can then compute the thermal resistance: $R_{\mathrm{th,gas}} = 1/(h_{\mathrm{eff}} A_{\mathrm{surface}})\approx 2.6\times 10^{7}~\mathrm{K/W}$, where $h_{\mathrm{eff}}$ is the heat transfer coefficient for the partial He atmosphere and $A_{\mathrm{surface}}\approx25.6~\upmu \mathrm{m^2}$ is the surface area of the device in contact with the gas. We can calculate the temperature rise considering only heat transfer to the gas for an electrical power of $P_{\mathrm{device}}\approx 4~\mathrm{nW}$ dissipated over a single device as $\Delta T_{\mathrm{gas}}\approx P_{\mathrm{device}}R_{\mathrm{th,gas}}\approx 0.1~\mathrm{K}$.

\subsubsection{Combined conduction and gas}
We can combine the thermal resistances from solid conduction ($R_{\mathrm{th,solid}}$) and gas conduction/convection ($R_{\mathrm{th,gas}}$) as thermal resistances to find the total thermal resistance, $R_{\mathrm{th}}=(R_{\mathrm{th,solid}}^{-1} + R_{\mathrm{th,gas}}^{-1})^{-1}\approx 2.6\times 10^{7}~\mathrm{K/W}$. Comparing the total thermal resistance ($R_{\mathrm{th}}$) to those for the solid and gas it is clear that the dominant heat transfer pathway is through the gas. Finally, we can compute the final temperature rise for an electrical power of $P_{\mathrm{device}}\approx 4~\mathrm{nW}$ dissipated over a single device: $\Delta T\approx P_{\mathrm{device}} R_{\mathrm{th}}\approx 0.1~\mathrm{K}$.  The calculated temperature rise of $\sim0.1$~K, confirms that Joule heating is negligible and that the observed spectral shifting is unlikely to be a thermal effect.

\subsection{Hole burning}
Spectral hole burning allows the instantaneous homogeneous linewidth of a single emitter to be probed, an indirect measurement of the true homogeneous linewidth. This method relies on saturation of the emitter where a pump laser is tuned to the centre of the spectral peak while a second probe laser is swept across the peak. With sufficient pump power the emitter will be in the saturation regime and additional power from the probe while detuned from the pump will result in additional signal as the probe laser excites the emitter at times when the slow spectral diffusion processes cause the homogeneous line to not be excited by the pump. As the probe laser is swept across the peak it samples a different temporal distribution than the pump laser until it is within the region defined by the convolution of two homogeneous lineshapes centred at the pump laser position. In this region the emitter is in saturation due to the pump laser and additional probe power causes minimal additional signal. This results in a saturation hole with a characteristic hole width $\Delta \nu_{\mathrm{hole}}$ given by~\cite{DeAbreu2023waveguide}:

\begin{equation}
    \Delta \nu_{\mathrm{hole}} = \Delta \nu_{\mathrm{hom}}\left(1+\sqrt{1+(P_{\mathrm{probe}}+P_{\mathrm{pump}})/P_{\mathrm{sat}}}\right)
    \label{eq:hole_burning}
\end{equation}

where $P_{\mathrm{probe}}$ is the probe laser power, $P_{\mathrm{pump}}$ is the pump laser power, $P_{\mathrm{sat}}$ is the saturation power, and $\Delta \nu_{\mathrm{hom}}$ is the homogeneous linewidth. The hole linewidth can then be used as a probe of the homogeneous linewidth where $\Delta \nu_{\mathrm{hom}}<\frac{1}{2}\Delta \nu_{\mathrm{hole}}$.

The hole burning measurement was performed by generating the pump and probe lasers from the single frequency of the tuneable laser using a fibre-coupled electro-optic phase modulator (EOM). The phase EOM was driven by applying the sum of two tones from a microwave generator (Zurich SHFSG) using a power combiner (Mini-Circuits ZFRSC-123-S+). This generates optical sidebands at the sum and difference frequencies of the two tones. Holeburning is performed by first detuning the laser from the spectral peak by an amount $\delta_0$. One of the microwave tones is fixed at the frequency $\delta_0$. The second microwave tone is swept across the detuning such that its frequency sweeps across the range $\Delta_s$, i.e. $\delta_0\pm\Delta_s/2$. The consequence of applying summed sinusoidal tones results in sum and difference frequencies. The impact of these is minimized by using a large detuning ($\sim$2~GHz) compared to the sweep range ($\sim$200~MHz) and ensuring that the modulation depth is set to minimize contributions from higher-order sidebands.

We perform hole burning on centre $A_3$ at $T=1.6$~K with optical powers from 20--400~nW for applied biases of 0~V and $-4$~V, where we observe a spectral shift of $-0.9$~GHz. Measurements at increased reverse biases were limited by poor SNR due to the decreased peak amplitude. We fit the model in equation~\ref{eq:hole_burning} to infer the low-power homogeneous linewidths ($\Delta\nu_{\mathrm{hom}}^{0}$) for the two bias points ($V_{\mathrm{a}}=0, -4~\mathrm{V}$) by measuring the hole linewidth ($\Delta\nu_{\mathrm{hole}}$) as the optical power ($P$) is reduced towards to the low-power limit:
\begin{equation}
    \Delta\nu_{\mathrm{hom}}^{0}\approx\lim\limits_{P \to 0} \Delta\nu_{\mathrm{hom}}(P,\mathrm{V_a})\lesssim 2~\Delta\nu_\mathrm{hole}
\end{equation}

This shows a clear relation which indicates that the observed hole widths exceed the low-power limit by approximately greater than or equal to a factor of two. The calculated homogeneous linewidths in the low-power limit are shown below along with their fitted saturation powers:

\begin{equation}
\Delta\nu_{\mathrm{hom}}^{0} \lesssim \begin{cases}
    31(5)~\mathrm{MHz} & \text{where }  \mathrm{V_a} = 0~\mathrm{V}~~~~,~P_{\mathrm{sat}}=14(8)~\mathrm{nW}\\
    37(8)~\mathrm{MHz} & \text{where } \mathrm{V_a} = -4~\mathrm{V}~,~P_{\mathrm{sat}}=24(19)~\mathrm{nW} 
\end{cases}
\end{equation}

\subsection{Comparison to bulk thermal shift and broadening}
\begin{figure*}[h]
  \makebox[\linewidth][c]{\includegraphics[width=180mm]{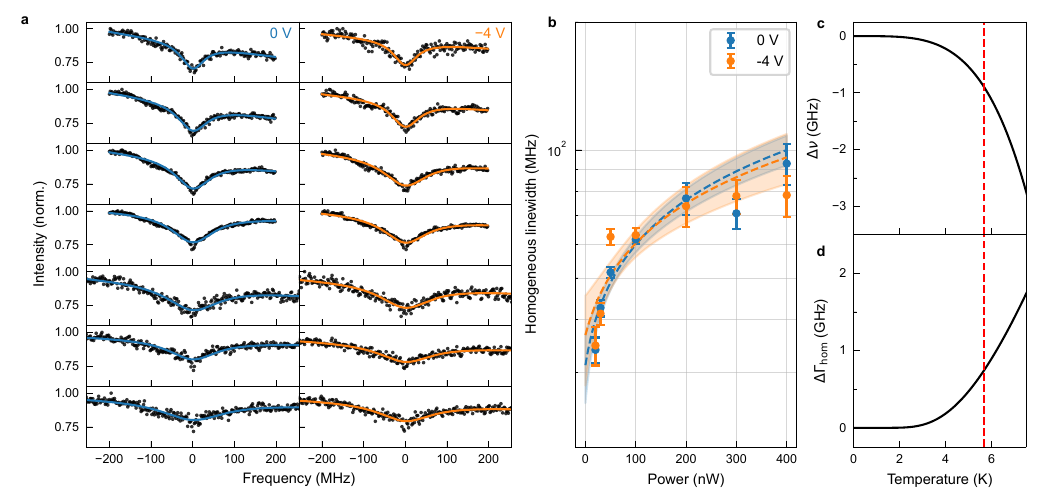}}%
  \caption{\textbf{Hole-burning. a}, Hole-burning measurements at 0~V (left) and $-4$~V (right) with powers of 20, 30, 50, 100, 200, 300, and 400~nW (top to bottom). \textbf{b}, Optical power dependence on homogeneous linewidth. \textbf{c}, SPL frequency shift vs. temperature following the bulk model from ref.~\cite{Bergeron2020}. \textbf{d}, Homogeneous broadening vs. temperature following the bulk model. The dashed red line at 5.7~K indicates the temperature required for the observed $-0.9$~GHz shift.}
  \label{fig:holeburning}
\end{figure*}

We can also compare the observed spectral behaviour to the thermal broadening and shifting models which have been measured for T centre ensembles in bulk $^{28}$Si~\cite{Bergeron2020}. Figure~\ref{fig:holeburning}c,d show these peak shift and broadening trends. The T centre's emission frequency relates to the system temperature by $\nu\propto A T^4$, where $A=-0.866~\mathrm{MHz}/T^4$ for T centres in $^{28}$Si~\cite{Bergeron2020}. Using this relation we calculate the temperature rise required for a $\Delta \nu=-0.9$~GHz shift to be $\Delta T\approx5.7$~K. Heating of the T centre can also be measured by broadening of the homogeneous line, which has been characterized in bulk spectroscopic measurements~\cite{Bergeron2020}. For the 5.7~K temperature rise required for a $-0.9$~GHz shift, this corresponds to broadening of $\Delta \Gamma=0.78$~GHz. 

This phenomenological broadening model includes both homogeneous and inhomogeneous sources. However, if we consider thermal broadening as predominantly contributing to homogeneous broadening this model provides an upper bound on the hole burning linewidth, $\Delta \nu_{\mathrm{hole}}>2(\Delta \nu_{\mathrm{hole}} + \Delta\Gamma)$. The lower limit of this arises when $\Delta \nu_{\mathrm{hole}}\to 0$ such that $\Delta \nu_{\mathrm{hole}}>1.6$~GHz. We find that even with a laser power of 400~nW and an applied bias of $-4$~V the hole burning linewidth does not exceed 200~MHz, confirming that the shifting is not due to thermal effects.

\clearpage
\section{Spin-dependent Transition Splitting modulation}
\label{supmat:gh_modulation}

\begin{figure}[h]
  \makebox[\linewidth][c]{\includegraphics{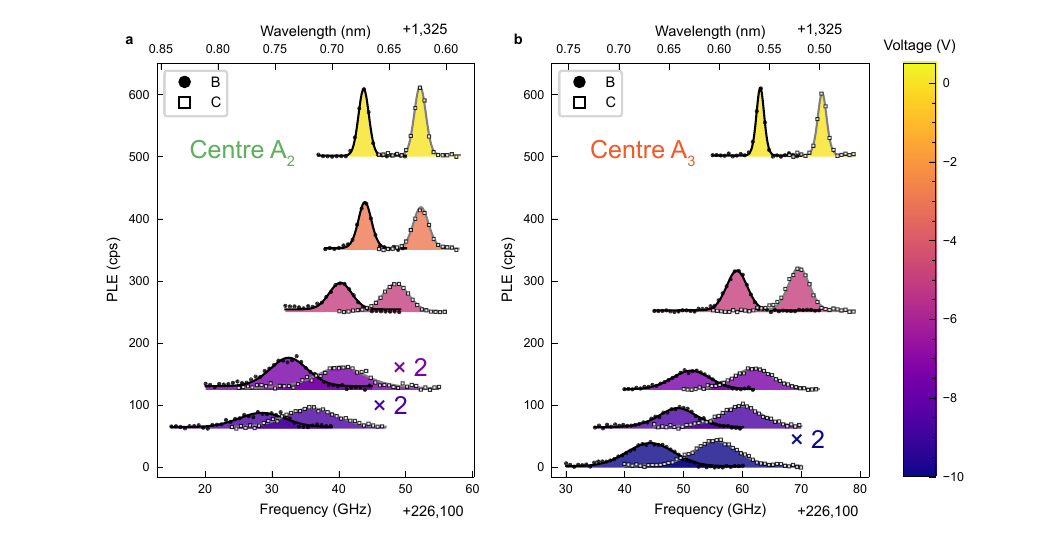}}%
  \caption{\textbf{Pump-probe PLE spectra with applied bias voltage.} Centres $A_2$ and $A_3$ were measured with a magnetic field $B=600$~mT, shown in (a) and (b), respectively. The data points for the B(C) transitions are shown with black circles (white squares). The spectra correspond to applied biases ($V_a$), offset by $50\cdot(V_a+10)$.}
  \label{fig:gh_mod_ple_spectra}
\end{figure}

\begin{figure}[h]
  \makebox[\linewidth][c]{\includegraphics{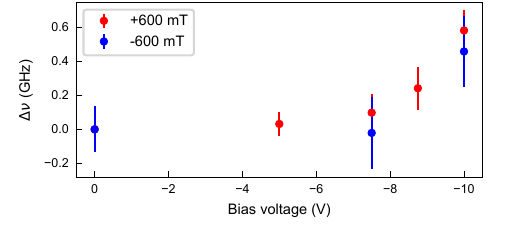}}%
  \caption{\textbf{Comparison of splitting modulation for centre $A_3$ with anti-parallel B-field directions.} The magnetic field aligned along the $[\bar{1}10]$ ($[1\bar{1}0]$) directions are shown in red (blue).}
  \label{fig:gh_mod_pos_neg}
\end{figure}

Figure~\ref{fig:gh_mod_ple_spectra} shows the PLE spectra and fits of the spin-dependent B and C transitions for centres $A_2$ and $A_3$ from which the results shown in Fig.~\ref{fig:gh_efield} in the main text were extracted. Centre $A_2$ was measured with applied biases of $0, -3, -5, -7.5, $ and $-8.75$~V. Centre $A_3$ was measured with applied biases of $0, -5, -7.5, -8.75$ and $-10$~V. Figure~\ref{fig:gh_mod_pos_neg} shows the splitting modulation measured with anti-parallel magnetic field directions, with the data for $+600$~mT being the same as presented in the main text.

\clearpage
\section{Stark Shift Response Time}
\label{supmat:pulsed_stark}

We measure the response time of the T centre Stark shift for centre $A_2$ using a pulsed electrical bias to characterize the performance of the integrated p-i-n diode and ensure that the T centre's emission frequency can be rapidly modulated for quantum operations. The measurement utilizes time-resolved PLE with a tuneable laser resonant with the T centre's ZPL, with $100$~ns optical pulses applied on a $4~\mu s$ period. The laser wavelength was set to 1325.675~nm to maximize the emission signal when the electrical pulse is applied during the optical pulse, this is detuned from the zero-electric-field peak of 1325.646~nm so that it does not excite the centre until it is Stark shifted.

\begin{figure}[h]
  \makebox[\linewidth][c]{\includegraphics[width=86mm]{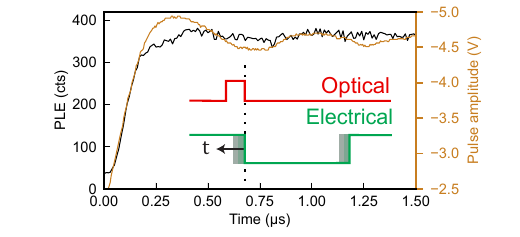}}%
  \caption{\textbf{Response to an electric pulse.} The electrical pulse (yellow) is swept backwards across the optical pulse. The PLE signal (black) shows the emission from the T centre increasing as the electrical pulse overlaps the optical pulse and causes a Stark shift.}
  \label{fig:pulsed_stark}
\end{figure}

A $2~\upmu$s electrical pulse is applied across the p-i-n diode by a voltage amplifier (Thorlabs HVA200) triggered with a digital delay generator with an output pulse amplitude of 1.5~V and a pulse offset voltage of 0.8~V. This pulse is attenuated by a $10\times$ voltage attenuator (due to the limited output voltage range) before being amplified with a gain of $-20\times$ resulting in a voltage step from $-1.6$~V to $-4.6$~V. The start of the electrical pulse is set to 100~ns following the optical pulse. This start time was swept backwards in 10~ns steps and the luminescence transient was collected from the T centre and recorded as a function of the electrical delay with 5~s integration time (Fig.~\ref{fig:pulsed_stark}). By scanning the delay we use the intensity as a probe of the emission frequency as it dynamically shifts across the Stark shift region. This approach is similar to procedures performed on other platforms~\cite{Cao2014Ultrafast}.

We find that the spectral shift exhibits a time response that tracks that electrical pulse. We measure the PLE signal rise time to be $t_{10-90\%}=160$~ns. As this measurement is a convolution of the true electrical RC-time of the device and the 100~ns duration of the optical pulse, this value represents an instrument-limited upper bound. This is comparable to the electrical response time, indicating that the diode characteristics and external electronics are the limiting factor to the Stark effect response time. This shows that Stark modulation can feasibly reach rates in excess of $5$~MHz in these devices, with the potential for significant improvements with optimized device designs.

\clearpage
\section{Observation of Shelving in a Dark Charge State}
\label{supmat:charge_state_shelving}

This section details an experimental investigation of the shelving dynamics of a single T centre into a dark charge state under the influence of an applied electric field. A combined optical and electrical pulsing scheme was used to induce shelving and observe emission from the centre following conversion back to the TX$_0$ state. This process is similar to exciton storage in quantum dots~\cite{Lundstrom1999Exciton, Giroday2011Exciton}. Correlation measurements confirm that the luminescence peaks before and after shelving are anti-correlated, corresponding to luminescence from a single T centre, as expected based on available decay pathways.

\begin{figure}[h]
  \makebox[\linewidth][c]{\includegraphics[width=180mm]{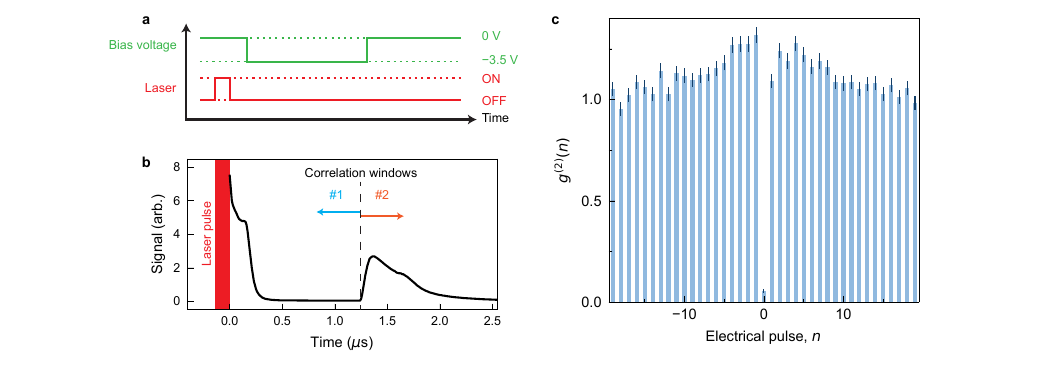}}%
  \caption{\textbf{Applying an electrical pulse following optical excitation causes shelving in a dark state. a}, Diagram of the pulse sequence for the laser and p-i-n diode bias voltage. \textbf{b}, Integrated intensity measurements with optical excitation followed by an electrical pulse. The start of the second peak aligns with the end of the electrical pulse. Two windows are used for correlation measurements, with the initial luminescence peak in window 1 and the delayed luminescence peak in window 2. \textbf{c}, Second-order correlation between the two peaks in (b) showing anti-correlation for $n=0$.}
  \label{fig:charge_state_shelving}
\end{figure}

Figure~\ref{fig:charge_state_shelving}a shows the pulse sequence used to observe shelving. In this sequence, centre $A_2$ was first optically excited to the TX$_0$ state. A short-duration reverse-bias electrical pulse was then applied to the p-i-n diode. Figure~\ref{fig:charge_state_shelving}b shows the histogram of photon arrival times, constructed by integrating the luminescence counts over many pulse cycles. The data shown in Fig.~\ref{fig:charge_state_shelving}b illustrates the shelving process:
\begin{enumerate}
    \item \textbf{Initial luminescence:} The first peak observed corresponds to the optical excitation of the T centre from its ground state to the excited TX$_0$ state, followed by radiative decay.
    \item \textbf{Non-radiative rate enhancement:} When the reverse-bias pulse begins, the initial luminescence peak is abruptly truncated, with the integrated luminescence dropping to the background level. This non-radiative rate enhancement causes shelving in the dark charge state.
    \item \textbf{Delayed luminescence:} After the bias voltage returns to 0~V, we observe a secondary luminescence peak. This represents re-population of the TX$_0$ excited state from the dark charge state, followed by radiative recombination.
\end{enumerate}

Figure~\ref{fig:charge_state_shelving}c shows the second-order autocorrelation measurement between the two luminescence peaks shown in Fig.~\ref{fig:charge_state_shelving}b. This measurement verifies that both of the luminescence peaks originate from the same emitter and confirms the mutual exclusivity of the decay pathways. We observed anti-correlation between the two measurement windows within a single pulse sequence, and no correlation between subsequent pulse sequences. For each excitation, a single T centre in the TX$_0$ state has two possible paths: it can either decay during the first window, radiatively or non-radiatively; or it can be shelved in the dark charge state and then decay during the second window (after the electric field is removed). A single centre cannot emit a photon in both windows for a single pulse cycle as optical excitation is possible only during the laser pulse, the reverse-bias electrical pulse cannot excite the centre, and following the electrical pulse the bias is returned to a level lower than that required for electroluminescence. 

\begin{figure}[h]
  \makebox[\linewidth][c]{\includegraphics[width=180mm]{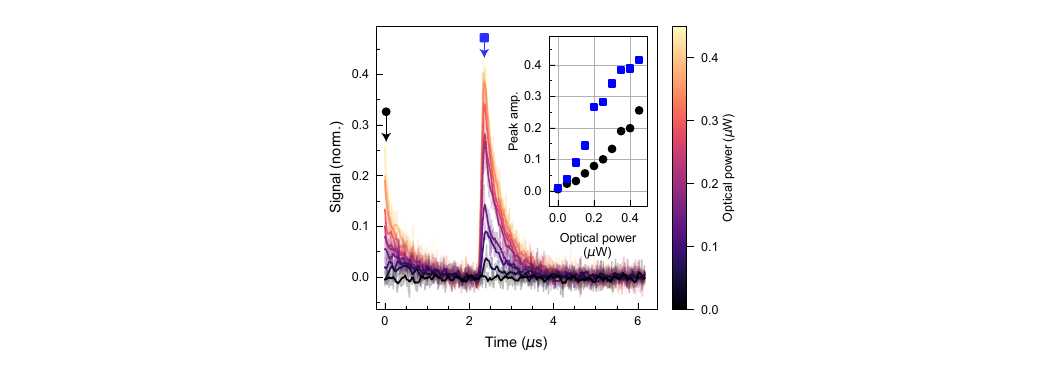}}%
  \caption{\textbf{Sweeping the optical power for a fixed electrical pulse.} The amplitudes of both luminescence peaks decreases as the optical power is reduced. The raw data is shown with partial transparency and the Savitsky-Golay filtered data (window=21, order=3) is shown as solid lines. The inset shows the peak amplitude of the filtered data for the initial (black circle) and delayed (blue square) luminescence peaks.}
  \label{fig:charge_state_shelving_power_sweep}
\end{figure}

\begin{figure}[h]
  \makebox[\linewidth][c]{\includegraphics[width=180mm]{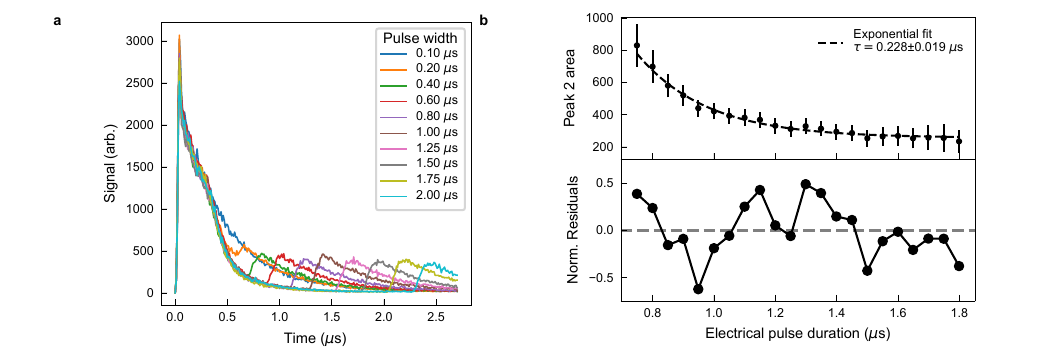}}%
  \caption{\textbf{Sweeping the electrical pulse width for a fixed optical power. a}, Integrated photon counts showing the initial and delayed luminescence peaks for varying electrical pulse widths. \textbf{b}, Area of the second peak (top panel) for electrical pulses from 0.8~$\upmu$s to 1.8~$\upmu$s. The data (black circles) are fitted to a single exponential (dashed line) with a time constant $\tau=228(19)$~ns corresponding to the dark state lifetime. Normalized residuals are plotted in the bottom panel.}
  \label{fig:charge_state_shelving_electrical_width_sweep}
\end{figure}

Figure~\ref{fig:charge_state_shelving_power_sweep} shows the two luminescence peaks for a sweep of the optical power. The possibility of the electrical transient causing re-excitation is rejected as measurements without optical excitation show only background luminescence.

Figure~\ref{fig:charge_state_shelving_electrical_width_sweep}a shows the two luminescence peaks for a sweep of the electrical pulse width. At short pulse durations we do not observe significant conversion to the dark state. As the pulse duration increases the second peak emerges, tracking the end of the electrical pulse. The luminescence decay is fitted to the sum of a bi-exponential decay (initial peak) and a rise-time limited exponential decay (equation~\ref{eq:double_decay}). Figure~\ref{fig:charge_state_shelving_electrical_width_sweep}b shows the second peak area is plotted for varying electrical pulse durations (top panel). The area of the second peak is calculated as $a_2(\tau_2^{\mathrm{fall}}-\tau_2^{\mathrm{rise}})$ and fitted to a single exponential for electrical pulse durations from 0.8~$\upmu$s to 1.8~$\upmu$s. The time constant of the fit corresponds to a dark state lifetime of 228(19)~ns.

\begin{equation}
\label{eq:double_decay}
    I(t) = a_1^{\mathrm{fast}}\exp\left(\frac{-(t-t_1^{\mathrm{fast}})}{\tau_1^{\mathrm{fast}}}\right) + a_1^{\mathrm{slow}}\exp\left(\frac{-(t-t_1^{\mathrm{slow}})}{\tau_1^{\mathrm{slow}}}\right)  + a_2\left[\exp\left(\frac{-(t-t_2)}{\tau_2^{\mathrm{fall}}}\right) - \exp\left(\frac{-(t-t_2)}{\tau_2^{\mathrm{rise}}}\right)\right]
\end{equation}

The data shown in Fig.~\ref{fig:charge_state_shelving}b was captured using an optical power of 0.4~$\upmu$W with the laser wavelength set to 1325.655~nm (detuned from the zero-electric-field point of 1325.646~nm to maximize the signal intensity). The total pulse sequence was 3~$\upmu$s with a 100~ns laser pulse and a 1~$\upmu$s electrical pulse (delayed 300~ns after the laser pulse). Electrical pulses were amplified to $-3.5$~V using a Thorlabs HVA200 high-voltage amplifier with a 1~MHz bandwidth, with the voltage returning to $0$~V following the pulse.

The data shown in Fig.~\ref{fig:charge_state_shelving_power_sweep} was captured using optical powers with the laser wavelength set to 1325.71~nm (detuned from the zero-electric-field point of 1325.646~nm to maximize the signal intensity). The total pulse sequence was 8~$\upmu$s with a 20~ns laser pulse and a 3~$\upmu$s electrical pulse (delayed 1~$\upmu$s after the laser pulse). Electrical pulses of $-1.4$~V were applied using an SRS DG645 delay generator, with the voltage returning to $0$~V following the pulse.

The data shown in Fig.~\ref{fig:charge_state_shelving_electrical_width_sweep}a was captured using optical powers with the laser wavelength set to 1325.645~nm. The total pulse sequence was 3~$\upmu$s with a 100~ns laser pulse and the electrical pulse delayed 300~ns after the laser pulse. Electrical pulses were amplified to $-1.0$~V using a Thorlabs HVA200 high-voltage amplifier with a 1~MHz bandwidth, with the voltage returning to $0$~V following the pulse.
\end{document}